\documentclass[twocolumn,superscriptaddress,amsmath,amssymb,floatfix,aps,notitlepage,nokeys,pre]{revtex4}
\usepackage{xr-hyper}
\usepackage{mathrsfs, hyperref}
\usepackage{amssymb, amsbsy, amsmath, latexsym, dsfont, array, layout, graphics,mathrsfs,braket,amsfonts,amsthm,
  amssymb,graphicx,subfigure, youngtab,color,bm,mathtools,braket,verbatim,url,cleveref,natbib,hypernat,extarrows,inputenc,multirow,bbm}
\usepackage{csquotes}
  
\usepackage[usernames,dvipsnames]{xcolor}
\usepackage{graphicx,psfrag}
\usepackage{amsfonts}
\usepackage[figuresright]{rotating}  
\usepackage{amssymb}
\usepackage{amsmath}
\usepackage{subfigure}
\usepackage{multirow}
\usepackage{tabularx}
\usepackage[resetlabels]{multibib}
\newcites{supp}{Supplementary References}
\usepackage{amssymb, amsbsy, amsmath, latexsym, dsfont, array, layout, graphics,mathrsfs,braket,amsfonts,amsthm,
  amssymb,graphicx,subfigure,dcolumn, youngtab,color,bm,mathtools,braket,verbatim,url,cleveref,natbib,hypernat,extarrows,inputenc,multirow}
\usepackage[usernames,dvipsnames]{xcolor}

\usepackage[mathlines]{lineno}% Enable numbering of text and display math

\allowdisplaybreaks
\begin{document}
\title{Omni-directional domain wall modes protected by fragile states}

\author{Pegah Azizi}
\affiliation{%
Department of Civil, Environmental, and Geo- Engineering,
University of Minnesota, Minneapolis, MN 55455, US}
\author{Siddhartha Sarkar}
\affiliation{%
Department of Physics, University of Michigan, Ann Arbor, MI 48109, USA
}
\author{Kai Sun}
\affiliation{%
Department of Physics, University of Michigan, Ann Arbor, MI 48109, USA
}
\author{ Stefano Gonella}%
\email{sgonella@umn.edu}
\affiliation{%
Department of Civil, Environmental, and Geo- Engineering,
University of Minnesota, Minneapolis, MN 55455, US}%
\date{Accepted 6 August 2024; published 20 August 2024; updated on arXiv 28 January 2025}
\text{Published DOI:\href{https://journals.aps.org/prb/abstract/10.1103/PhysRevB.110.L060102}{10.1103/PhysRevB.110.L060102}}

\begin{abstract}
So-called fragile topological states of matter challenge our conventional notion of topology by lacking the robustness typically associated with topological protection, thereby displaying elusive manifestations that are difficult to harness for wave control. In this Letter, we leverage the recent discovery of fragile topological states in special classes of structural kagome lattices to document the availability of domain wall elastic wave modes that are directly traceable to fragile topology and, yet, exhibit remarkably strong signatures that support omni-directionality. We design twisted kagome bi-domains comprising two topologically distinct sublattices - one trivial and the other fragile topological - sharing a common bandgap and meeting at a domain wall. The two phases are achieved via carefully engineered surface cut patterns that modify the band landscape of the underlying lattices in complementary fashions, leading to dichotomous irreps landscapes. Under these circumstances, a domain wall-bound mode emerges within the shared bandgap and displays remarkable stability against domain wall orientation and introduced defects. We corroborate these findings via $\mathbf{k}\cdot\mathbf{p}$ Hamiltonian and Jackiw-Rabbi analysis and validate them experimentally through laser vibrometry tests on a prototype endowed with water jet-induced perforation patterns.
\end{abstract}

\maketitle
Maxwell lattices~\cite{Maxwell_1864,CALLADINE_1978} have received substantial attention in recent years due to an array of unique mechanical properties, including the ability to localize deformation at floppy edges~\cite{Mao_Lubensky_maxwell_topo_2018,Mao_Lubensky_APS_2011,Kane_Lubensky_Nphys_2014,Lubensky_et_all_2015,Rocklin_et_all_natcom_2017} and focus stress at domain walls (DWs), thus eliciting protection against fracture and buckling~\cite{Zhang-Mao_Fracture_NJP_2018,Paulose-et-al_Topo-Buckling_PNAS_2015}, and tunable multistability~\cite{Xiu_et_all_bistable_pnas_2022}. A popular subclass of Maxwell systems, kagome lattices have been extensively studied, especially in the realm of ideal configurations characterized by perfect hinges, 
for the peculiar properties rooted in their topology and symmetries~\cite{Sun_et_all_pnas_2012,Danawe_et_all_cornermode_PhysRevB_2021,Nassar-et-al_Microtwist_JMPS_2020,Chen_et_all_kagome_PhysRevB_2018,Riva_et_all_topo_kagome_appliedphys_2018,Schaeffer_Ruzzene_kagome_appliedphys_2015,Bertoldi_et_all_natrevmats_2017}. Many topological properties of Maxwell lattices have been shown to be preserved, albeit diluted, in transitioning from \textit{ideal} to \textit{structural} lattices that can be physically fabricated, where the bonds can carry flexural deformation and the hinges are replaced by internal clamps or by ligaments featuring finite stiffness against rotations~\cite{Azizi_et_all_PhysRevLett_2023,Charara_et_all_PhysRevApplied_2021,WIDSTRAND_et_all_IJSS_2023,CHEN_et_all_IJSS_2022,Jihong_et_all_PhysRevLett_2018,Zhang_et_all_kagome_2023_PhysRevApplied}.

Recent studies have put forth the notion of fragile topology~\cite{Po_et_ell_fragile_PhysRevLett_2018,Bradlyn_et_all_fragile_PhysRevB_2019,Ahn_et_all_fragile_PhysRevX_2019,Bouhon_et_all_fragile_PhysRevB_2019,Wieder_Bernevig_fragile_arxiv_2018,Song_et_all_fragile_science_2020,Peri_et_all_fragile_science_2020,wieder_et_all_fragile_natcom_2020,Lian_et_all_fragile_PhysRevB_2020,Bouhon_et_all_PhysRevB_2020,de_Paz_et_all_fragile_PhysRevResearch_2019,Peri_et_all_fragile_PhysRevLett_2021}. Fragile bands qualify as topological in that they do not admit a symmetric, exponentially localized Wannier representation that preserves all the symmetries of the system. However, unlike stable topology, their obstruction to the atomic limit can be eliminated by coupling with fully topologically trivial bands. Additionally, these states generally lack a well defined conventional bulk-edge correspondence~\cite{Xu_et_all_catalogue_arxiv_2022,Nur_et_all_PhysRevLett_2020,Peng_et_all_natcom_2022,Kooi_et_all_npjQuantumMaterials_2021,PRB2020_ROBUST}. A recent investigation by Azizi et al. explored fragile topology in the context of twisted kagome lattices in the so-called self-dual configuration~\cite{Azizi_et_all_PhysRevLett_2023}. The term \textit{self-dual}, introduced by Fruchart et al. within their treatment of mechanical duality of twisted kagome lattices~\cite{Fruchart_et_all_duality_nat_2020} refers to the critical point in configuration space that separates families of dual pairs, which feature identical phononic properties~\cite{Fruchart_et_all_duality_nat_2020,Danawe_et_all_cornermode_PhysRevB_2021,Danawe_et_all_PhysRevLett_2022}. The authors of~\cite{Azizi_et_all_PhysRevLett_2023} proved theoretically and demonstrated experimentally the existence of fragile topological states in self-dual kagome lattices in the \textit{structural} configuration featuring ligament-like hinges at the lattice sites. However, while they identified a number of localized modes, for none of them a direct connection to fragile topology or claims of topological protection could be made.

This Letter aims at filling this conceptual gap by deliberately searching for localized modes that are unequivocally linked to, and protected by, the fragile topology. We seek such modes along the DWs of bi-domain lattices obtained by adjoining two structural self-dual twisted kagome (SSTK) sublattices with different topological character - one trivial and the other fragile topological - making sure that the two phases are designed to be spectrally compatible. The proposed approach draws partial inspiration from the behavior of mechanical analogs of the quantum valley-Hall effect (QVHE)~\cite{Marino_et_all_QVHE_PhysRevX_2015,Raj_Ruzzene_QVHE_NewJournalPhys_2017,Jihong_et_all_QVHE_PhysRevApplied_2019,Liu_Semperlotti_QVHE_PhysRevApplied_2019,Qian_et_all_QVHE_PhysRevB_2018}, where domain-bound modes can be generated at the DW of two subdomains characterized by opposite valley topologies bounding a shared bandgap (BG)~\cite{Pan_et_all_QVHE_PhysRevB_2015}. A by-product objective of our study is to address a limitation of QVHE analogs, which support DW modes along various orientations~\cite{CHENvalley2019,Wu2017valleyNatCom} except for specific forbidden directions. In contrast, here we put forth a strategy to achieve truly omni-directional DW mode localization capability along any arbitrary direction. 
%By-product objective of our study is precisely to overcome limitation of QVHE analogs, which only work for systems with low intervalley scattering and for DWs oriented along selected directions\pa{~\cite{CHENvalley2019}}. In contrast to QVHE, we document a substantial robustness of the observed modes against different DW orientations and the introduction of defects. 

\begin{figure}[t!]
\includegraphics[width=\columnwidth]{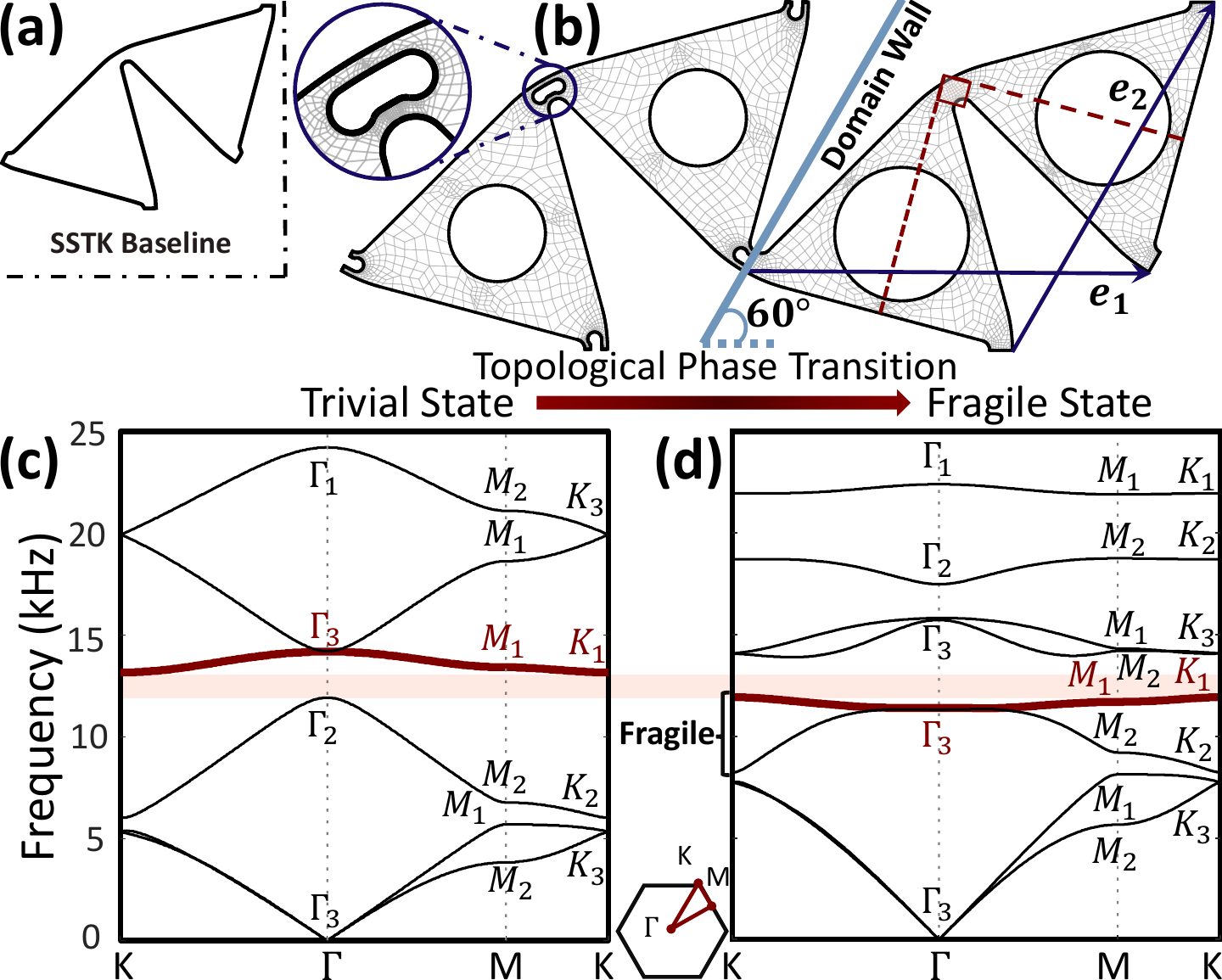}
\caption{(a) Unit cell of SSTK baseline configuration. (b) Bi-domain lattice with DW obtained stitching a trivial and a fragile topological SSTK phases. Zoomed-in detail of the bean-shaped hole at the hinge of the trivial phase is presented in the inset. Both configurations feature large holes to modulate inertia and control frequency spectra. (c-d) Band diagrams of the trivial and fragile topological states, respectively. The thick maroon branch denotes the band undergoing inversion, leading to the closing and reopening of the shared BG (shaded in pink) marking the topological phase transition. The irreps at HSPs of the BZ confirm the topological nature of the third and fourth bands in (d).}\label{fig:model}
\end{figure}
\begin{figure}[t!]
\includegraphics[width=\columnwidth]{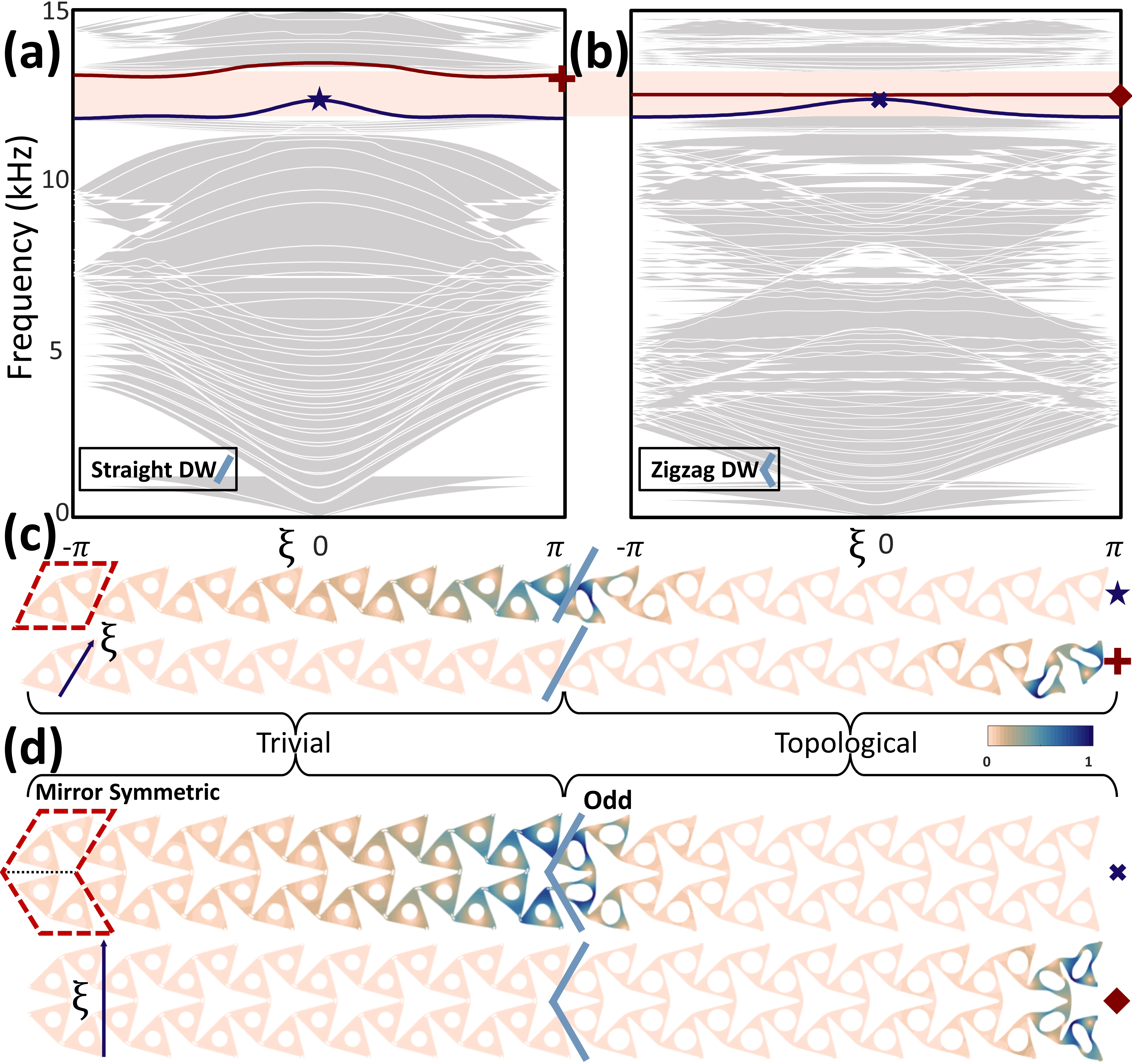}
\caption{\label{fig:sup}(a-b) Band diagrams for two differently oriented 16-cell SSTK supercells (with unit cells highlighted) forming (a) $60^{\circ}-$DW and (b) $90^{\circ}-$ZGDW, with branches for the DW mode and edge modes highlighted in blue and maroon, respectively. (c-d) Wave functions (mode shapes) of localized DW and edge states for the $60^{\circ}-$DW and $90^{\circ}-$ZGDW, sampled at $\xi=0$ and $\xi=\pi$. The left and right sides consist of trivial and topological states, respectively.} 
 \end{figure}
\begin{figure*}[t]
\includegraphics[width=\textwidth]{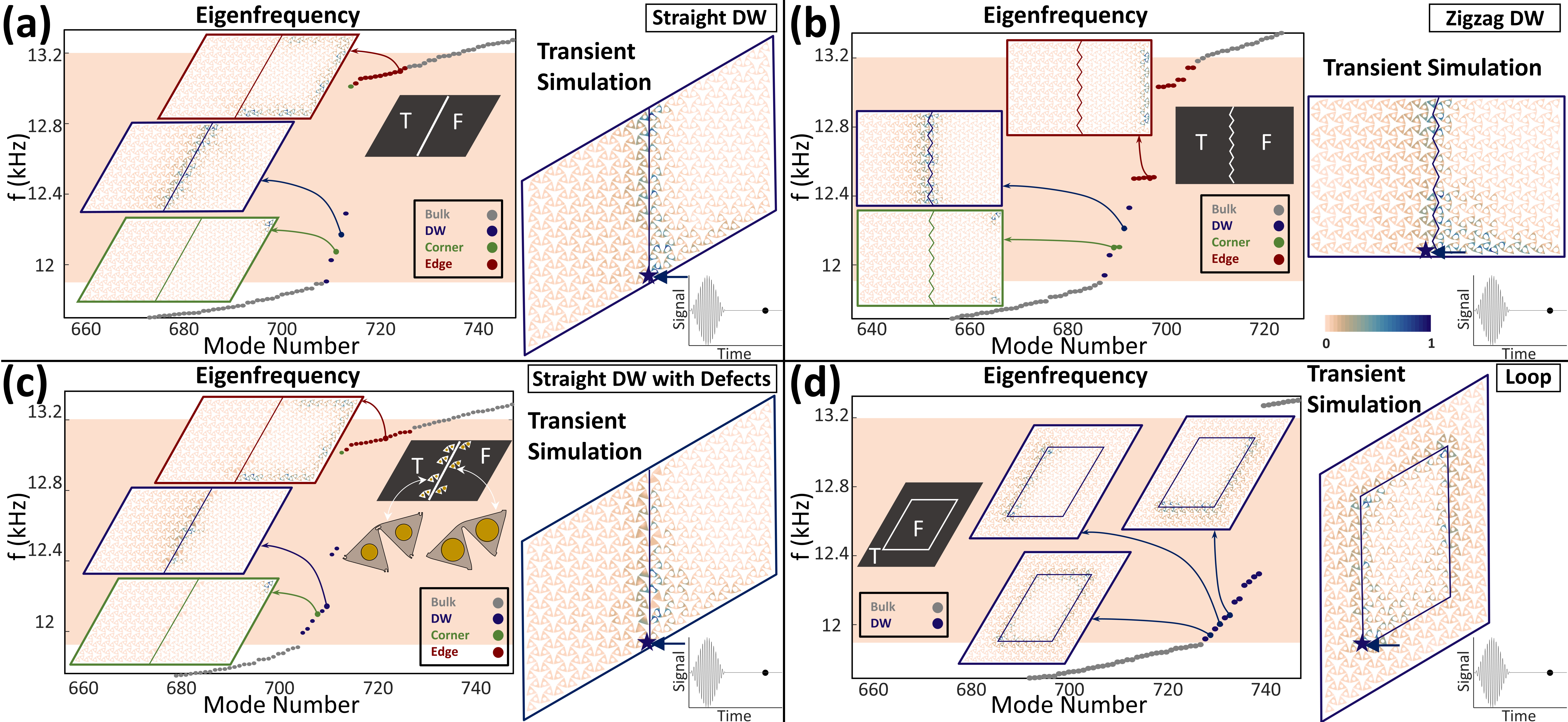}
\caption{(a-f) Dynamical response for six scenarios: (a) $60^{\circ}-$DW, (b) $90^{\circ}-$ZGDW, (c) $20^{\circ}-$ZGDW, (d) $40^{\circ}-$ZGDW, (e) $60^{\circ}-$DW with randomly located defects, and (f) internal loop. For each case, the left panel contains eigenfrequencies in the frequency interval of the shared BG, color-coded to distinguish bulk modes and different types of localized modes, including DW (with the corresponding frequency indicated), corner, and edge states, with three sample eigenfields shown in the insets. The right panel depicts snapshots of propagating wavefields (from full-scale transient simulations) induced by a narrow-band burst force excitation, polarized perpendicularly to the DW. In all cases, the wave is unequivocally localized along the DW.}
\label{fig:sim}
\end{figure*}

We begin our analysis resuming the SSTK configuration presented in~\cite{Azizi_et_all_PhysRevLett_2023}, which features fragile topological bands. Recall that the configuration consists of a self-dual kagome in which the triangles are solid patches, taken to be thin to realize plane-stress conditions, connected by finite-thickness ligaments. Our objective is to extract from this baseline geometry two topologically distinct derivative configurations - one trivial and the other topological - that are simultaneously geometrically and spectrally compatible. Geometric compatibility requires that the lattices can be seamlessly stitched to form a DW, which implies that they shall share the same unit cell primitive lattice vectors and $C_{3v}$ point group symmetry. Moreover, in anticipation of the construction of an experimental prototype, we seek configurations that are practically feasible under the constraints of the available fabrication methods. Specifically, we want the lattices to be mono-material, as to be agilely obtainable via subtractive manufacturing (e.g., machining or water-jet cutting) from a slab of homogeneous, low-damping material with uniform and predictable properties, e.g., a metal. We also require that the two lattices deviate from each other only in terms of different patterns of cuts introduced a posteriori as a perturbation of the shared baseline SSTK geometry. This ensures that the thickness of the hinges, which is a critical parameter, is preserved across the DW. Finally, we want the two configurations to be spectrally compatible, i.e., to feature a shared full BG in the mid-frequency region of the spectrum.

The first step of the design process consists of modifying the topological SSTK baseline configuration (Fig.~\ref{fig:model}(a)) to induce a transition to a trivial configuration, marked by the closing and reopening of the mid-frequency BG. An avenue towards this phase transition is by softening the hinge ligament of the unit cell as explored in~\cite{Azizi_et_all_PhysRevLett_2023}. In light of the above-stated mono-material constraint, here we seek an analogous softening effect purely by virtue of geometry. To this end, we introduce isolated cuts or clusters of cuts in the neighborhood of the hinges. After a few design iterations (supplemental material (SM), \ref{sec.1}), we eventually land on the bean-shaped hole shown in the inset of Fig.~\ref{fig:model}(b). Notably, this addition results in the desired topological band inversion along with a substantial increase in BG width (SM.\ref{sec.1}). To ensure spectral compatibility between the new trivial (with hole) and the original topological (hole-less) configurations, we modulate their effective mass densities by endowing the unit cells of both configurations with additional holes and we fine tune their relative diameters until the two BGs have comparable width and align in the desired interval of the spectrum. The resulting geometries are shown in Fig.~\ref{fig:model}(b) in their stitched configuration, with the trivial phase featuring hinge holes and small density-correcting holes on the left of the DW, and the topological phase featuring intact hinges and large density-correcting holes on the right. The primitive lattice vectors are denoted as $\mathbf{e}_1$ and $\mathbf{e}_2$, with $\mathbf{e}_2$ aligned parallel to the DW, forming a $60^{\circ}$ angle with the horizontal. The corresponding band diagrams are shown in Figs.~\ref{fig:model}(c-d), where we can observe the above-mentioned band inversion at the $\Gamma$ point, whereby the fourth band (highlighted in maroon) appears above and below the shared BG, respectively. Additionally, we report irreducible representations (irreps) calculated at the high symmetry points (HSPs) of the Brillouin zone (BZ) following the procedure and notation given in the Bilbao crystallography server~\cite{aroyo_et_all_Billcrys_BulgChemCommun_2011,Aroyo_et_all_Billcrys_2006I,Aroyo_et_all_Billcrys_2006II,vergniory_et_all_Billcrys_Phys.Rev.E_2017,elcoro_et_all_Billcrys_Journal_of_Applied_Crystallography_2017} based on the recently developed method of topological quantum chemistry, which classifies all topological states that are protected by spatial symmetries~\cite{bradlyn_et_all_TQC_Nat_2017,cano_et_all_TQC_Phys.Rev.B_2018}. The irreps, which encapsulate the symmetry properties of the eigenmodes at the HSPs, provide a recipe to qualify the topology of isolated bands. The two-dimensional (2D) irrep $\Gamma_3$ and 1D irrep $\Gamma_2$ undergo band inversion from the case of Figs.~\ref{fig:model}(c) and ~\ref{fig:model}(d), resulting in bands 3 and 4 for having now HSP irreps $\Gamma_3-M_1\oplus M_2-K_1\oplus K_2$ in Fig.~\ref{fig:model}(d), which implies these two bands are indeed fragile topological, protected by $C_3$ symmetry~\cite{Song_et_all_fragile_science_2020,Azizi_et_all_PhysRevLett_2023}.

\begin{figure*}[t]
\includegraphics[width=\textwidth]{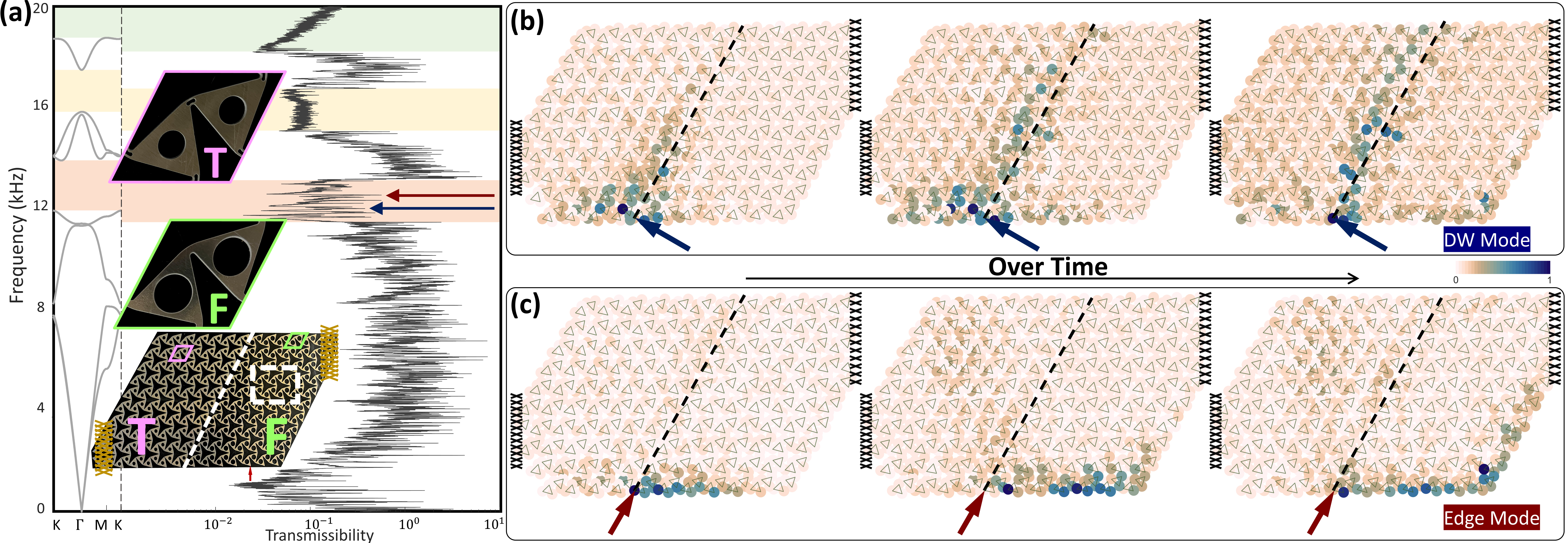}
\caption{\label{fig:exp}(a) Laser vibrometry-acquired experimental transmissibility curve with highlighted attenuation regions reasonably matching the BG frequencies of the fragile phase spectrum plotted on the left. Insets provide details of the trivial (T) and fragile (F) unit cells of the water jet-cut specimen. White dashed box denotes a region where response is sampled. Blue and maroon arrows mark the excitation frequencies for the DW and edge modes activation, respectively. (b-c) Snapshots of the measured wavefields induced by 30-cycle burst with carrier frequencies corresponding to the DW and edge modes. Force polarization is indicated by the arrows. Color bar indicates velocity amplitude normalized by the highest value in each frame.}
\end{figure*}

Having established a DW across which a topological phase transition occurs, we proceed to ask whether such DW can be the site of strongly localized modes and whether such modes enjoy any protection that is attributable to the fragile topology. Here, we consider two distinct versions of the bi-domain lattice that can be realized by stitching trivial and fragile phases, one featuring a $60^{\circ}-$DW and the other $90^{\circ}-$zigzag DW (ZGDW). For each, we perform supercell analysis using a dedicated supercell. Specifically, the first supercell (Fig.~\ref{fig:sup}(c)) is obtained from the tessellation of a single SSTK unit cell and intercepts the $60^{\circ}-$DW already discussed in Fig.~\ref{fig:model}. The other (Fig.~\ref{fig:sup}(d)) consists of the tessellation of a mirror symmetric macrocell encompassing two unit cells, and results in a $90^{\circ}-$ZGDW globally perpendicular to the supercell itself and parallel to the direction of tessellation. The corresponding supercell band diagrams are shown in Figs.~\ref{fig:sup}(a-b), respectively. In both configurations, we identify two relatively flat modes in the BG, which we recognize, from the mode shapes depicted in Figs.~\ref{fig:sup}(c-d), as a DW mode and an edge mode, color-coded in blue and maroon, respectively. The DW mode presents a few peculiarities. First, the availability of DW modes for both $60^{\circ}-$DW and $90^{\circ}-$ZGDW (as well as potentially along any other arbitrary direction, as demonstrated later in the Letter) represents an increase in versatility compared to QVHE, where DW modes are not achievable for angles of $90^\circ$ (and equivalently $30^\circ$ and $150^\circ$) (SM.\ref{sec.2}). Here, the emergence of DW  modes for any angle can be understood from the isotropic  nature of the $\Gamma$ point in a $C_3$ symmetric system. %For instance, the availability of DW modes for both \pa{$60^{\circ}-$DW} and \pa{$90^{\circ}-$ZGDW} (and \pa{potentially} along any other arbitrary direction, \pa{as documented later in the Letter}) represents an increase in flexibility with respect to the case of QVHE, where \pa{DW modes are not achievable for $90^\circ$ and equivalently $30^\circ$, and $150^\circ$ angles from one of the lattice vectors (SM, Sec.~2.~\cite{sup})}. 
Moreover, most conventional topologically protected DW branches span the BG with quasi-constant finite slopes, denoting modes that propagate at finite velocities for any carrier frequency in the BG~\cite{Raj_Ruzzene_QVHE_NewJournalPhys_2017,Jihong_et_all_QVHE_PhysRevApplied_2019}. Here, in contrast, the relatively flat band results in the ability of the lattice to localize the signal in space with pronounced persistence over a frequency subinterval of the BG, while maintaining the response overall gapped. As for the edge modes, it is worth noting that they appear only in the fragile phase, where they display a high decay rate.

The emergence of the DW modes resulting from the topological phase transition marked by the band inversion between the $\Gamma_2$ and $\Gamma_3$ irreps, can be analytically confirmed using Jackiw-Rebbi (JR) analysis~\cite{jackiwRebbi1976}. In SM.\ref{sec.2}, starting from a $\mathbf{k}\cdot\mathbf{p}$ Hamiltonian~\cite{dresselhaus_et_all_group_2007} near the $\Gamma$ point of the BZ and considering only the bands with $\Gamma_2$ and $\Gamma_3$ irreps, we theoretically prove that DW modes must emerge between the trivial and fragile phases for any DW direction. Furthermore, we prove that the DW mode shapes obtained from the supercell analysis, shown in Figs.~\ref{fig:sup}(c-d), are indeed caused by the topological phase transition between the trivial and fragile states by comparing them against those obtained from the JR analysis. It is worthwhile mentioning that, since the $90^{\circ}-$ZGDW is mirror symmetric, a DW mode at this interface would have to be even or odd under the mirror reflection. It can be seen from Fig.~\ref{fig:sup}(d) that the DW mode has indeed odd parity under the mirror- a feature captured by JR analysis. On the other hand, the $60^{\circ}-$DW breaks the mirror symmetry, hence the DW mode does not have a definite parity, also predicted by the JR analysis.
 
Next, we evaluate eigenfrequencies and mode shapes for six finite bi-domain configurations depicted in Fig.~\ref{fig:sim}, including $60^{\circ}-$DW (Fig.~\ref{fig:sim}(a)), $90^{\circ}-$ZGDW (Fig.~\ref{fig:sim}(b)), $20^{\circ}-$ZGDW (Fig.~\ref{fig:sim}(c)), $40^{\circ}-$ZGDW (Fig.~\ref{fig:sim}(d)), $60^{\circ}-$DW with defects (Fig.~\ref{fig:sim}(e)), and loop (Fig.~\ref{fig:sim}(f)). In the configuration of Fig.~\ref{fig:sim}(e), we randomly select seven unit cells close to the DW and we fill their central holes with inclusions, choosing a density ratio $\rho_{inc}=0.8\rho_{bulk}$ to perturb the lattice order. In each of the cases in Figs.~\ref{fig:sim}(a-e), we report localized states in the BG frequency range: we recognize DW, corner, and edge modes, and we highlight a few randomly selected ones for each type by showing the corresponding eigenfunctions. Notably, in Figs.~\ref{fig:sim}(a-e), edge and corner modes localize exclusively along the edges of the fragile phase, consistent with the findings from supercell analysis. Finally, in the case of Fig.~\ref{fig:sim}(f), we construct a closed DW loop to confine a parallelogram of fragile phase within a larger domain of the trivial phase and we perform a similar inference of the localized modes.  

Subsequently, we provide additional evidence of DW emergence in the considered scenarios via full-scale transient simulations. Snapshots of computed propagating wavefields are displayed in the right part of each panel of Figs.~\ref{fig:sim}(a-f) under a point-force excitation, polarized perpendicular to the DW, prescribed at the bottom tip of the DW in the form of a 17-cycle narrow-band tone burst, with carrier frequency set at $\sim$12.2 kHz. Notably, in all cases, the wavefields exhibit pronounced localization along the DW, regardless of the DW orientation and the presence of defects. 

We proceed to validate our theoretical findings through laser vibrometry experiments performed on a physical prototype (for detailed information regarding the fabrication, vibrometer specifications, and experimental setup refer to SM.\ref{sec.3}). The specimen, along with zoomed-in details of the fragile and trivial unit cells, are displayed in the insets of Fig.~\ref{fig:exp}(a). Our initial objective is to experimentally determine the BG frequencies. To accomplish this, we prescribe a broadband pseudorandom vertical excitation at a bottom edge point of the fragile phase (marked by a red arrow), we measure the in-plane vertical velocity at scan points inside a sampling region located away from the DW and the edges (dashed box in the inset of Fig.~\ref{fig:exp}(a)), and we normalize the average velocity values with those recorded at the excitation point to construct a transmissibility curve versus frequency, as presented in Fig.~\ref{fig:exp}(a) ( See SM.\ref{sec.4} for analogous curves obtained via full-scale steady-state simulation). Notably, the transmissibility reveals three distinct regions of attenuation, highlighted in lighter shades, which align well, within the expected deviations between model and experiments, with the BGs derived from Bloch analysis. The region shaded in pink marks the mid-frequency BG of interest, within which we carry out two separate transient tests to empirically document the emergence of DW and edge modes. To this end, we excite the bottom tip of the DW with bursts with carrier frequencies marked by the blue and maroon arrows in Fig.~\ref{fig:exp}(a). We note that the frequency discrepancies between model and experiments prevent us from being able to know precisely the frequency of the localized modes a priori. Similarly, the transmissibility plot does not carry a sufficient spectral signature of them that can be leveraged. As a result, the identification of the carriers requires a heuristic approach in which we probe the frequency interval expected to be of interest based on analogy with simulations by conducting a series of tests with varying carriers until we detect the emergence of the desired modes. Eventually, we identify 12 kHz and 12.5 kHz as the two frequencies of interest for the DW and edge modes, respectively. Specifically, if we apply a 30-cycle burst with a frequency of $\sim$12 kHz, polarized perpendicular to the DW direction, we generate a wave that propagates along the DW, three snapshots of which are depicted in Fig.\ref{fig:exp}(b). Similarly, Fig.~\ref{fig:exp}(c) presents three snapshots of a wave induced by an excitation with a carrier frequency of $\sim$12.5 kHz, polarized parallel to the DW, which features localization along the edge (of the fragile phase only, consistent with the eigenmode prediction). In summary, in all cases we record strong localization that persists even deep into the propagation process, before reflections from the opposite boundary begin to interfere with the incident wave to form a standing pattern.

P.A. acknowledges the support of the UMN CSE Graduate Fellowship. S.G. acknowledges support from the National Science Foundation (grant CMMI-2027000). S.S. and K.S. acknowledge the support from the Office of Naval Research (grant MURI N00014-20-1-2479).

%\bibliographystyle{apsrev4-1}
%\bibliography{ref}% Produces the bibliography via BibTeX.

%\input{refbbl.bbl}
%\bibliographystyle{naturemag}
%\bibliography{ref}% Produces the bibliography via BibTeX.
\bibliographystyle{apsrev4-1}
\bibliography{ref}% Produces the bibliography via BibT

\onecolumngrid

%%%%%%%%%%%%%%%%%%%%%%%%%%%%%%%%%%%%%%
%%   Supplementary Information
%%%%%%%%%%%%%%%%%%%%%%%%%%%%%%%%%%%%%%
\section*{\Large\bf Supplemental Material}
\counterwithout{figure}{section} 
\makeatletter
\renewcommand \thesection{S-\@arabic\c@section}
\renewcommand\thetable{S\@arabic\c@table}
\renewcommand{\thefigure}{S\arabic{figure}}
\renewcommand \theequation{S\@arabic\c@equation}
\makeatother

\maketitle
\section{Design Process for Deriving Trivial and Fragile Phases from the Structural Self-Dual Twisted Kagome (SSTK) Baseline and Topological Phonon Band Inversion}\label{sec.1}

\begin{figure}[h!]
   \centering
  \includegraphics[width = 0.9\textwidth]{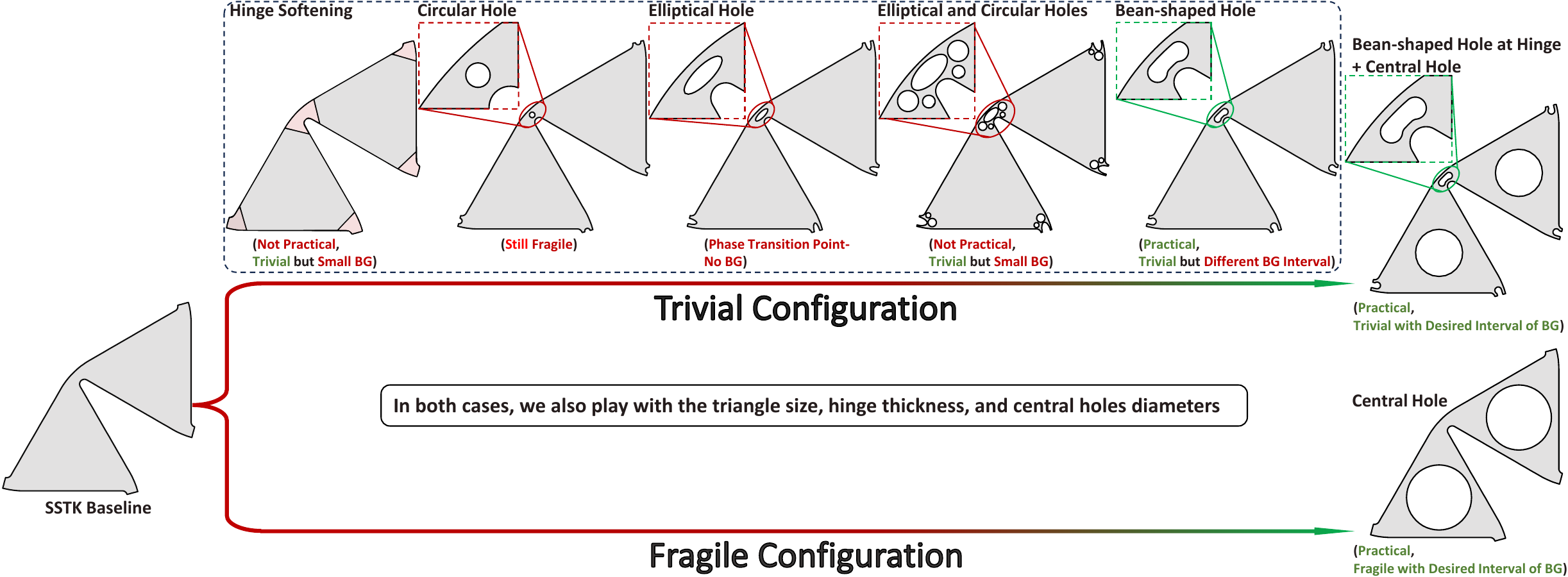}
   \caption{Unit cell of SSTK baseline configuration, along with the details of several design iterations considered to derive two spectrally compatible unit cells featuring trivial (top row) and fragile (bottom row) states. In each step, any cut patterns at the hinge, if applicable, are magnified in the inset. Practical and not practical refer to  configurations that can and cannot be realized via conventional fabrication methods available to us, respectively.}
   \label{fig:des}
\end{figure}

In this section, we report the design thought process involved in deriving two spectrally compatible unit cells featuring distinct topological character. The starting point for this derivation is the SSTK baseline configuration presented in~\cite{Azizi_et_all_PhysRevLett_2023}, which is characterized by fragile bands. Sequential steps of the design process are depicted in Fig.~\ref{fig:des}. The upper row of the figure illustrates several iterations that lead to the finalization of the desired trivial unit cell, featuring a bean-shaped hole at the hinge and inertia-correcting holes at the center of the unit cell. The inset in each step magnifies the details of the cut patterns at the hinge. The lower row displays the modified SSTK baseline configuration endowed with central inertia-correcting holes leading calibrated to yield a mid-frequency bandgap (BG) in the same frequency range of the trivial unit cell. 

\begin{figure}[h!]
   \centering
 \includegraphics[width = 0.9\textwidth]{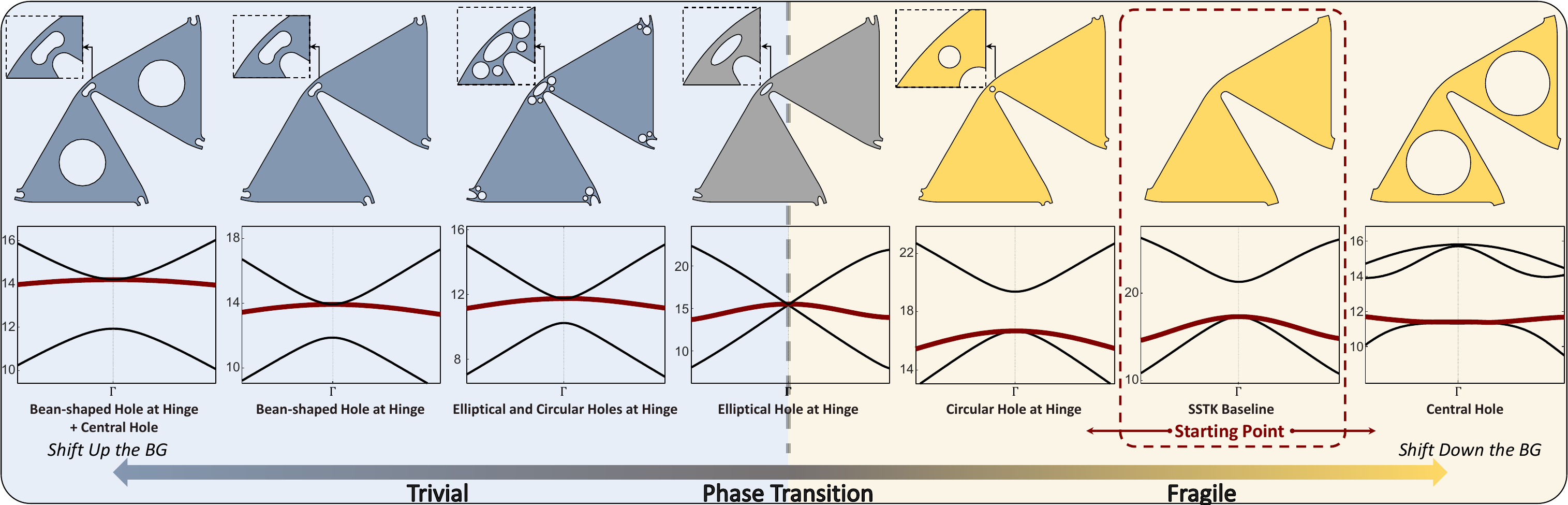}
 \caption{Topological phase transition diagram of SSTK unit cells with various cut patterns. The corresponding band diagrams near $\Gamma$ are shown for each step, with magnified insets depicting any cut patterns at the hinge if applicable. The fourth band (highlighted in maroon) undergoes band inversion during the transition.}
   \label{fig:BD}
\end{figure}

Additionally, we illustrate the evolution of the band diagram near the $\Gamma$ point for each design step, organized from the trivial phase to the fragile configuration in Fig.~\ref{fig:BD}. The dispersion curves reveal phonon band inversion of the fourth mode (highlighted in maroon) during the transition from trivial to fragile states. Notably, the BG closes at the configuration where an elliptical hole is introduced at the hinge.

\section{\texorpdfstring{$\mathbf{k}\cdot\mathbf{p}$}{Lg} Hamiltonian and Jackiw-Rebbi (JR) Analysis of the domain wall (DW) mode}\label{sec.2}
In the main text, we mention that the band inversion between the 2-fold degenerate $\Gamma_3$ (2D irreducible representation, bands 4-5 in Fig.~1(c)) and the singly degenerate $\Gamma_2$ (1D irreducible representation, band 3 in Fig.~1(c)) bands from the trivial phase give rise to the fragile set of bands (bands 3-4 in Fig.~1(d)) in the fragile phase (see also~\cite{Azizi_et_all_PhysRevLett_2023} for more details). Here, using JR analysis~\cite{jackiwRebbi1976}, we show the existence of exponentially localized DW modes for any arbitrary oriented DW, including $60^\circ-$DW and $90^\circ-$zigzag DW (ZGDW) between the trivial and the fragile phases due to the band inversion.
\subsection{The \texorpdfstring{$\mathbf{k}\cdot\mathbf{p}$}{Lg} Hamiltonian}
To begin, recall that the eigenmodes we consider are the eigenmodes of the Navier-Cauchy equation $( \boldsymbol{\nabla} (\lambda+\mu)\boldsymbol{\nabla}\cdot+ \boldsymbol{\nabla}\cdot\mu\boldsymbol{\nabla})\mathbf{u}(\mathbf{r}) = -\rho \omega^2\mathbf{u}(\mathbf{r})$, where $\lambda$ and $\mu$ are the Lam\'e coefficients of the elastic material, $\rho$ is density, $\boldsymbol{\nabla} = \hat{\mathbf{x}}\partial_x+\hat{\mathbf{y}}\partial_y$ is the spatial gradient operator, and $(\cdot)$ stands for the dot product of vectors. For a periodic system (system with discrete translation symmetry), due to Bloch's theorem the eigenmodes $\mathbf{u}(\mathbf{r})$ are of the form $\mathbf{u}_\mathbf{k}(\mathbf{r}) = \tilde{\mathbf{u}}_\mathbf{k}(\mathbf{r}) e^{i\mathbf{k}\cdot\mathbf{r}}$, where $\tilde{\mathbf{u}}_\mathbf{k}(\mathbf{r})$ is a periodic funciton and $\mathbf{k}$ is the wave-vector. Plugging this ansatz into the Navier-Cauchy equation, we get the effective equation $H(\mathbf{k})\tilde{\mathbf{u}}_\mathbf{k}(\mathbf{r}) \equiv -( (\boldsymbol{\nabla}+i\mathbf{k})\frac{\lambda+\mu}{\rho}(\boldsymbol{\nabla}+i\mathbf{k})\cdot+ (\boldsymbol{\nabla}+i\mathbf{k})\cdot \frac{\mu}{\rho} (\boldsymbol{\nabla}+i\mathbf{k}))\tilde{\mathbf{u}}_\mathbf{k}(\mathbf{r}) =\omega_\mathbf{k}^2\tilde{\mathbf{u}}_\mathbf{k}(\mathbf{r})$, where $H(\mathbf{k})$ is the Hamiltonian operator ($H(\mathbf{k})$ is a Hermitian operator). The band structure is the plot of eigenvalues $\omega_\mathbf{k}$ as a function of $\mathbf{k}$.
\begin{figure}[h!]
    \centering
    \includegraphics[width = 0.5\textwidth]{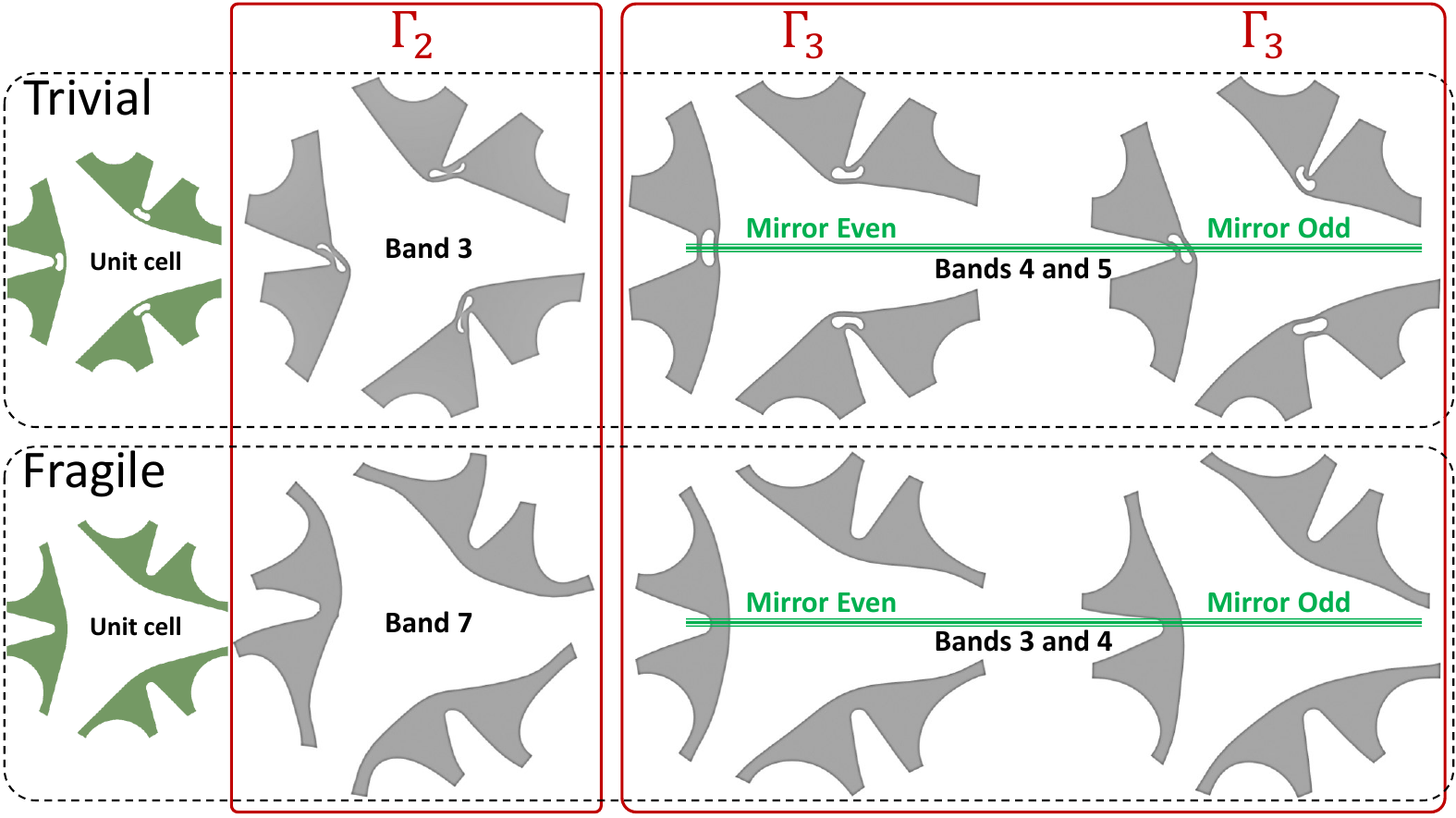}
    \caption{Wigner-Seitz unit cells of the trivial and fragile states along with the mode shapes correspond to $\Gamma_2$ and $\Gamma_3$ representations. The modes in the first $\Gamma_3$ column are even ($p_x$-type) under horizontal mirror $\hat{M}_y$, whereas the modes in the second $\Gamma_3$ column are odd ($p_y$ type) under $\hat{M}_y$.}
    \label{fig:symmod}
\end{figure}

Next, we note that the point group symmetry of the system, $C_{3v}$, consists of three fold ($120^\circ$) rotation operation $\hat{C}_3$ (the $240^\circ$ rotation is $\hat{C}_3^2$) and mirror $\hat{M}_y$ which flips the $y$-coordinate $y\rightarrow -y$ (the other two mirrors are $\hat{C}_3 \hat{M}_y \hat{C}_3^{-1}$, $\hat{C}_3^2 \hat{M}_y \hat{C}_3^{-2}$). The $\Gamma_3$ modes $\tilde{\mathbf{u}}_1^{(3)}(\mathbf{r})$ and $\tilde{\mathbf{u}}_2^{(3)}(\mathbf{r})$ transform like components of a 2D vector (or $p_x$-$p_y$ orbitals) under $\hat{C}_3$ and $\hat{M}_y$ (see Fig.~\ref{fig:symmod})
% \begin{equation}
\begin{align}
    &(\hat{C}_3 \tilde{\mathbf{u}}_1^{(3)})(\mathbf{r}) \equiv \mathbf{R}(2\pi/3)\tilde{\mathbf{u}}_1^{(3)}(\mathbf{R}^{-1}(2\pi/3)\mathbf{r}) = +\cos(2\pi/3)\tilde{\mathbf{u}}_1^{(3)}(\mathbf{r})+ \sin(2\pi/3)\tilde{\mathbf{u}}_2^{(3)}(\mathbf{r}) = -\frac{1}{2}\tilde{\mathbf{u}}_1^{(3)}(\mathbf{r})+ \frac{\sqrt{3}}{2}\tilde{\mathbf{u}}_2^{(3)}(\mathbf{r}),\nonumber\\
    &(\hat{C}_3 \tilde{\mathbf{u}}_2^{(3)})(\mathbf{r}) \equiv \mathbf{R}(2\pi/3)\tilde{\mathbf{u}}_2^{(3)}(\mathbf{R}^{-1}(2\pi/3)\mathbf{r}) = -\sin(2\pi/3)\tilde{\mathbf{u}}_1^{(3)}(\mathbf{r})+ \cos(2\pi/3)\tilde{\mathbf{u}}_2^{(3)}(\mathbf{r}) = -\frac{\sqrt{3}}{2}\tilde{\mathbf{u}}_1^{(3)}(\mathbf{r})- \frac{1}{2}\tilde{\mathbf{u}}_2^{(3)}(\mathbf{r}),\nonumber\\
    &\hspace{5cm}(\hat{M}_y \tilde{\mathbf{u}}_1^{(3)})(\mathbf{r}) \equiv\sigma_z\tilde{\mathbf{u}}_1^{(3)}(\sigma_z^{-1}\mathbf{r}) = \tilde{\mathbf{u}}_1^{(3)}(\mathbf{r}),\nonumber\\
    &\hspace{5cm}(\hat{M}_y \tilde{\mathbf{u}}_2^{(3)})(\mathbf{r}) \equiv \sigma_z\tilde{\mathbf{u}}_2^{(3)}(\sigma_z^{-1}\mathbf{r}) = -\tilde{\mathbf{u}}_2^{(3)}(\mathbf{r}),
\end{align}
% \end{equation}
where $\mathbf{R}(2\pi/3)$ is the rotation matrix that rotates a vector by angle $2\pi/3$, $\sigma_z = \text{Diag}\{1,-1\}$ is the Pauli matrix which flips the sign of $y$ component of a vector. On the other hand, the mode $\tilde{\mathbf{u}}^{(2)}(\mathbf{r})$ corresponds to $\Gamma_2$ representation is odd under all mirrors, but remains invariant under $\hat{C}_3$ (see Fig.~\ref{fig:symmod})
\begin{equation}
\begin{gathered}
    (\hat{C}_3 \tilde{\mathbf{u}}^{(2)})(\mathbf{r}) \equiv \mathbf{R}(2\pi/3)\tilde{\mathbf{u}}^{(2)}(\mathbf{R}^{-1}(2\pi/3)\mathbf{r}) = \tilde{\mathbf{u}}^{(2)}(\mathbf{r}),\\
    (\hat{M}_y \tilde{\mathbf{u}}^{(2)})(\mathbf{r}) \equiv\sigma_z\tilde{\mathbf{u}}^{(2)}(\sigma_z^{-1}\mathbf{r}) = -\tilde{\mathbf{u}}^{(2)}(\mathbf{r}).
\end{gathered}
\end{equation}
These equations can be succinctly written as
\begin{equation}
\label{eq:transformations}
\begin{gathered}
     \{(\hat{C}_3 \tilde{\mathbf{u}}_1^{(3)}),(\hat{C}_3 \tilde{\mathbf{u}}_2^{(3)}),(\hat{C}_3 \tilde{\mathbf{u}}^{(2)})\}(\mathbf{r}) = \{\tilde{\mathbf{u}}_1^{(3)}, \tilde{\mathbf{u}}_2^{(3)},\tilde{\mathbf{u}}^{(2)}\}(\mathbf{r}) \rho(\hat{C}_3),\text{ where } 
     \rho(\hat{C}_3) =  \begin{pmatrix}-\frac{1}{2} & -\frac{\sqrt{3}}{2} & 0\\
     \frac{\sqrt{3}}{2} & -\frac{1}{2} & 0\\
     0 & 0 & 1\end{pmatrix},\\
     \{(\hat{M}_y \tilde{\mathbf{u}}_1^{(3)}),(\hat{M}_y \tilde{\mathbf{u}}_2^{(3)}),(\hat{M}_y \tilde{\mathbf{u}}^{(2)})\}(\mathbf{r}) = \{\tilde{\mathbf{u}}_1^{(3)}, \tilde{\mathbf{u}}_2^{(3)},\tilde{\mathbf{u}}^{(2)}\}(\mathbf{r}) \rho(\hat{M}_y),\text{ where } 
     \rho(\hat{M}_y) =  \begin{pmatrix}1 & 0 & 0\\
     0 & -1 & 0\\
     0 & 0 & -1\end{pmatrix}.
\end{gathered}
\end{equation}
The matrices $\rho(\hat{C}_3)$ and $\rho(\hat{M}_y)$ are called the representation matrices of the corresponding symmetries. Furthermore, the complex conjugate of the Hamiltonian $H(\mathbf{k})$ at $\mathbf{k}$ is the Hamiltonian at $-\mathbf{k}$: $H^*(\mathbf{k}) = H(-\mathbf{k})$. Hence, $\omega_\mathbf{k}^2 \tilde{\mathbf{u}}_\mathbf{k}^*(\mathbf{r}) = (\omega_\mathbf{k}^2)^*\tilde{\mathbf{u}}_\mathbf{k}^*(\mathbf{r}) = H^*(\mathbf{k})\tilde{\mathbf{u}}_\mathbf{k}^*(\mathbf{r}) = H(-\mathbf{k})\tilde{\mathbf{u}}_\mathbf{k}^*(\mathbf{r})$ (we use the fact that the eigenfrequencies are real: $\omega_\mathbf{k}=\omega_\mathbf{k}^*$), which implies $\tilde{\mathbf{u}}_{-\mathbf{k}}(\mathbf{r})\propto\tilde{\mathbf{u}}_\mathbf{k}^*(\mathbf{r})$. From this, it is easy to see that at $\Gamma$ point ($\mathbf{k} = \mathbf{0}$), the eigenfunctions can be chosen as real. Hence, we take
\begin{equation}
    \{(\tilde{\mathbf{u}}_1^{(3)})^*,(\tilde{\mathbf{u}}_2^{(3)})^*,(\tilde{\mathbf{u}}^{(2)})^*\}(\mathbf{r}) = \{\tilde{\mathbf{u}}_1^{(3)}, \tilde{\mathbf{u}}_2^{(3)},\tilde{\mathbf{u}}^{(2)}\}(\mathbf{r}).
\end{equation}
To do an effective analysis near the $\Gamma$ point ($\mathbf{k} = \mathbf{0}$) including just the three bands, we do a projection of the full Hamiltonian $H(\mathbf{k})$ onto the three bands of importance. The projected Hamiltonian $h(\mathbf{k})$ has the form
\begin{equation}
\label{eq:Hprojection}
\begin{split}
    h_{\alpha\beta}(\mathbf{k}) &= \langle \tilde{\mathbf{u}}_\alpha | H(\mathbf{k})|\tilde{\mathbf{u}}_\beta\rangle, \text{ where } \tilde{\mathbf{u}}_\alpha(\mathbf{r}) \in \{\tilde{\mathbf{u}}_1^{(3)}(\mathbf{r}), \tilde{\mathbf{u}}_2^{(3)}(\mathbf{r}),\tilde{\mathbf{u}}^{(2)}(\mathbf{r})\}\\
    &= \int_\text{unit cell} d^2\mathbf{r}\, \tilde{\mathbf{u}}_\alpha^*(\mathbf{r})\cdot H(\mathbf{k})\tilde{\mathbf{u}}_\beta(\mathbf{r})\\
    &= \int_\text{unit cell} d^2\mathbf{r}\, \tilde{\mathbf{u}}_\alpha^*(\mathbf{r})\cdot ( -(\boldsymbol{\nabla}+i\mathbf{k})\frac{\lambda}{\rho}(\boldsymbol{\nabla}+i\mathbf{k})\cdot - (\boldsymbol{\nabla}+i\mathbf{k})\cdot \frac{\mu}{\rho} (\boldsymbol{\nabla}+i\mathbf{k}))\tilde{\mathbf{u}}_\beta(\mathbf{r})\\
    & = \int_\text{unit cell} d^2\mathbf{r}\, \frac{\lambda+\mu}{\rho} ((\boldsymbol{\nabla}+i\mathbf{k})\cdot\tilde{\mathbf{u}}_\alpha(\mathbf{r}))^* (\boldsymbol{\nabla}+i\mathbf{k})\cdot\tilde{\mathbf{u}}_{\beta}(\mathbf{r}) + \frac{\mu}{\rho} ((\boldsymbol{\nabla}+i\mathbf{k})\tilde{\mathbf{u}}_{\alpha}(\mathbf{r}))^* \boldsymbol{:}(\boldsymbol{\nabla}+i\mathbf{k})\tilde{\mathbf{u}}_{\beta}(\mathbf{r})\\
    & \hspace{3cm} \text{where } ((\boldsymbol{\nabla}+i\mathbf{k})\tilde{\mathbf{u}}_{\alpha}(\mathbf{r}))^* \boldsymbol{:}(\boldsymbol{\nabla}+i\mathbf{k})\tilde{\mathbf{u}}_{\beta}(\mathbf{r}) = \sum_{i,j \in \{x,y\}}((\partial_i+ik_i)\tilde{u}_{\alpha,j}(\mathbf{r}))^*(\partial_i+ik_i)\tilde{u}_{\beta,j}(\mathbf{r}),
\end{split}
\end{equation}
where from third to fourth line, we use integration by parts; since the functions $\mathbf{u}_\alpha(\mathbf{k})$ are periodic, the boundary terms in by parts integration are zero. Note that at $\mathbf{k} = \mathbf{0}$, the projected Hamiltonian $h(\mathbf{k} = \mathbf{0})$ is a diagonal matrix with diagonal entries being the squares of the eigenfrequencies of the three modes: $h(\mathbf{k} = \mathbf{0}) = \text{Diag}\{\omega_{\Gamma_3}^2,\omega_{\Gamma_3}^2,\omega_{\Gamma_2}^2\}$. Due to $\hat{C}_3$, the projected Hamiltonian satisfies the following identity
% \begin{subequations}
\begin{align}
    &h_{\alpha\beta}(\mathbf{R}(2\pi/3)\mathbf{k})\nonumber\\
    &= \int_\text{unit cell} d^2\mathbf{r}\, \frac{\lambda+\mu}{\rho} ((\boldsymbol{\nabla}+i \mathbf{R}(2\pi/3)\mathbf{k})\cdot\tilde{\mathbf{u}}_\alpha(\mathbf{r}))^* (\boldsymbol{\nabla}+i\mathbf{R}(2\pi/3)\mathbf{k})\cdot\tilde{\mathbf{u}}_{\beta}(\mathbf{r})\nonumber\\
    &\hspace{2.2cm}+ \frac{\mu}{\rho} ((\boldsymbol{\nabla}+i\mathbf{R}(2\pi/3)\mathbf{k})\tilde{\mathbf{u}}_{\alpha}(\mathbf{r}))^* \boldsymbol{:}(\boldsymbol{\nabla}+i\mathbf{R}(2\pi/3)\mathbf{k})\tilde{\mathbf{u}}_{\beta}(\mathbf{r})\nonumber\\
    &= \int_\text{unit cell} d^2\mathbf{r}\, \frac{\lambda+\mu}{\rho} ((\mathbf{R}(2\pi/3)(\mathbf{R}^{-1}(2\pi/3)\boldsymbol{\nabla}+i \mathbf{k}))\cdot\tilde{\mathbf{u}}_\alpha(\mathbf{r}))^* (\mathbf{R}(2\pi/3)(\mathbf{R}^{-1}(2\pi/3)\boldsymbol{\nabla}+i\mathbf{k}))\cdot\tilde{\mathbf{u}}_{\beta}(\mathbf{r})\nonumber\\
    &\hspace{2.2cm}+ \frac{\mu}{\rho} ((\mathbf{R}(2\pi/3)(\mathbf{R}^{-1}(2\pi/3)\boldsymbol{\nabla}+i\mathbf{k}))\tilde{\mathbf{u}}_{\alpha}(\mathbf{r}))^* \boldsymbol{:}(\mathbf{R}(2\pi/3)(\mathbf{R}^{-1}(2\pi/3)\boldsymbol{\nabla}+i\mathbf{k}))\tilde{\mathbf{u}}_{\beta}(\mathbf{r})\nonumber\\
    &= \int_\text{unit cell} d^2\mathbf{r}\, \frac{\lambda+\mu}{\rho} ((\mathbf{R}^{-1}(2\pi/3)\boldsymbol{\nabla}+i \mathbf{k})\cdot\mathbf{R}^{-1}(2\pi/3)\tilde{\mathbf{u}}_\alpha(\mathbf{r}))^* (\mathbf{R}^{-1}(2\pi/3)\boldsymbol{\nabla}+i\mathbf{k})\cdot\mathbf{R}^{-1}(2\pi/3)\tilde{\mathbf{u}}_{\beta}(\mathbf{r})\nonumber\\
    &\hspace{2.2cm}+ \frac{\mu}{\rho} ((\mathbf{R}^{-1}(2\pi/3)\boldsymbol{\nabla}+i\mathbf{k})\mathbf{R}^{-1}(2\pi/3)\tilde{\mathbf{u}}_{\alpha}(\mathbf{r}))^* \boldsymbol{:}(\mathbf{R}^{-1}(2\pi/3)\boldsymbol{\nabla}+i\mathbf{k})\mathbf{R}^{-1}(2\pi/3)\tilde{\mathbf{u}}_{\beta}(\mathbf{r})\nonumber\\
    &= \int_\text{unit cell} d^2\mathbf{r}'\, \frac{\lambda+\mu}{\rho} ((\boldsymbol{\nabla}'+i \mathbf{k})\cdot\mathbf{R}^{-1}(2\pi/3)\tilde{\mathbf{u}}_\alpha(\mathbf{R}(2\pi/3)\mathbf{r}'))^* (\boldsymbol{\nabla}'+i\mathbf{k})\cdot\mathbf{R}^{-1}(2\pi/3)\tilde{\mathbf{u}}_{\beta}(\mathbf{R}(2\pi/3)\mathbf{r}')\nonumber\\
    &\hspace{2.2cm}+ \frac{\mu}{\rho} ((\boldsymbol{\nabla}+i\mathbf{k})\mathbf{R}^{-1}(2\pi/3)\tilde{\mathbf{u}}_{\alpha}(\mathbf{R}(2\pi/3)\mathbf{r}'))^* \boldsymbol{:}(\boldsymbol{\nabla}+i\mathbf{k})\mathbf{R}^{-1}(2\pi/3)\tilde{\mathbf{u}}_{\beta}(\mathbf{R}(2\pi/3)\mathbf{r}') \text{ with }\mathbf{r}' = \mathbf{R}^{-1}(2\pi/3)\mathbf{r}\nonumber\\
    &= \int_\text{unit cell} d^2\mathbf{r}'\, \frac{\lambda+\mu}{\rho} ((\boldsymbol{\nabla}'+i \mathbf{k})\cdot \sum_\gamma\tilde{\mathbf{u}}_\gamma(\mathbf{r}')(\rho^{-1}(\hat{C}_3))_{\gamma\alpha})^* (\boldsymbol{\nabla}'+i\mathbf{k})\cdot\sum_\delta\tilde{\mathbf{u}}_\delta(\mathbf{r}')(\rho^{-1}(\hat{C}_3))_{\delta\beta}\nonumber\\
    &\hspace{2.2cm}+ \frac{\mu}{\rho} ((\boldsymbol{\nabla}'+i\mathbf{k})\sum_\gamma\tilde{\mathbf{u}}_\gamma(\mathbf{r}')(\rho^{-1}(\hat{C}_3))_{\gamma\alpha})^* \boldsymbol{:}(\boldsymbol{\nabla}'+i\mathbf{k})\sum_\delta\tilde{\mathbf{u}}_\delta(\mathbf{r}')(\rho^{-1}(\hat{C}_3))_{\delta\beta} \text{ using Eq.~\eqref{eq:transformations}}\nonumber\\
    &= \int_\text{unit cell} d^2\mathbf{r}'\, \frac{\lambda+\mu}{\rho} ((\boldsymbol{\nabla}'+i \mathbf{k})\cdot \sum_\gamma\tilde{\mathbf{u}}_\gamma(\mathbf{r}')(\rho^{T}(\hat{C}_3))_{\gamma\alpha})^* (\boldsymbol{\nabla}'+i\mathbf{k})\cdot\sum_\delta\tilde{\mathbf{u}}_\delta(\mathbf{r}')(\rho^{T}(\hat{C}_3))_{\delta\beta}\nonumber\\
    &\hspace{2.2cm}+ \frac{\mu}{\rho} ((\boldsymbol{\nabla}'+i\mathbf{k})\sum_\gamma\tilde{\mathbf{u}}_\gamma(\mathbf{r}')(\rho^{T}(\hat{C}_3))_{\gamma\alpha})^* \boldsymbol{:}(\boldsymbol{\nabla}'+i\mathbf{k})\sum_\delta\tilde{\mathbf{u}}_\delta(\mathbf{r}')(\rho^{T}(\hat{C}_3))_{\delta\beta} \text{ since }\rho^{-1}(\hat{C}_3) = \rho^{T}(\hat{C}_3)\nonumber\\
    &= \int_\text{unit cell} d^2\mathbf{r}'\, \frac{\lambda+\mu}{\rho} ((\boldsymbol{\nabla}'+i \mathbf{k})\cdot \sum_\gamma\tilde{\mathbf{u}}_\gamma(\mathbf{r}')\rho(\hat{C}_3)_{\alpha\gamma})^* (\boldsymbol{\nabla}'+i\mathbf{k})\cdot\sum_\delta\tilde{\mathbf{u}}_\delta(\mathbf{r}')(\rho^T(\hat{C}_3))_{\delta\beta}\nonumber\\
    &\hspace{2.2cm}+ \frac{\mu}{\rho} ((\boldsymbol{\nabla}'+i\mathbf{k})\sum_\gamma\tilde{\mathbf{u}}_\gamma(\mathbf{r}')\rho(\hat{C}_3)_{\alpha\gamma})^* \boldsymbol{:}(\boldsymbol{\nabla}'+i\mathbf{k})\sum_\delta\tilde{\mathbf{u}}_\delta(\mathbf{r}')(\rho^T(\hat{C}_3))_{\delta\beta}\nonumber\\
    &= \sum_{\gamma,\delta}\rho(\hat{C}_3)_{\alpha\gamma}\Big[\int_\text{unit cell} d^2\mathbf{r}'\, \frac{\lambda+\mu}{\rho} ((\boldsymbol{\nabla}'+i \mathbf{k})\cdot \tilde{\mathbf{u}}_\gamma(\mathbf{r}'))^* (\boldsymbol{\nabla}'+i\mathbf{k})\cdot\tilde{\mathbf{u}}_\delta(\mathbf{r}')\nonumber\\
    &\hspace{4.2cm}+ \frac{\mu}{\rho} ((\boldsymbol{\nabla}'+i\mathbf{k})\tilde{\mathbf{u}}_\gamma(\mathbf{r}'))^* \boldsymbol{:}(\boldsymbol{\nabla}'+i\mathbf{k})\tilde{\mathbf{u}}_\delta(\mathbf{r}')\Big](\rho^T(\hat{C}_3))_{\delta\beta},\nonumber\\
    &=\sum_{\gamma,\delta}\rho(\hat{C}_3)_{\alpha\gamma} h_{\gamma\delta}(\mathbf{k})(\rho^T(\hat{C}_3))_{\delta\beta}\nonumber\\
    \Rightarrow& h(\mathbf{R}(2\pi/3)\mathbf{k}) = \rho(\hat{C}_3) h(\mathbf{k})  \rho^T(\hat{C}_3).
\end{align}
% \end{subequations}
A similar calculation using $\hat{M}_y$ shows
\begin{equation}
    h(\sigma_z\mathbf{k}) = \rho(\hat{M}_y) h(\mathbf{k})  \rho^T(\hat{M}_y).
\end{equation}
Furthermore, using the fact that $H^*(\mathbf{k}) = H(-\mathbf{k})$, it can be seen that
\begin{equation}
    h^*(\mathbf{k}) = h(-\mathbf{k}).
\end{equation}
Together, the projected Hamiltonian satisfies the following properties
\begin{equation}
\label{eq:hSymmetries}
\begin{gathered}
    h(\mathbf{R}(2\pi/3)\mathbf{k}) = \rho(\hat{C}_3) h(\mathbf{k})  \rho^T(\hat{C}_3),\\
    h(\sigma_z\mathbf{k}) = \rho(\hat{M}_y) h(\mathbf{k})  \rho^T(\hat{M}_y),\\
    h^*(\mathbf{k}) = h(-\mathbf{k}).
\end{gathered}
\end{equation}
\begin{figure}[h!]
    \centering
    \includegraphics[width = 0.9\textwidth]{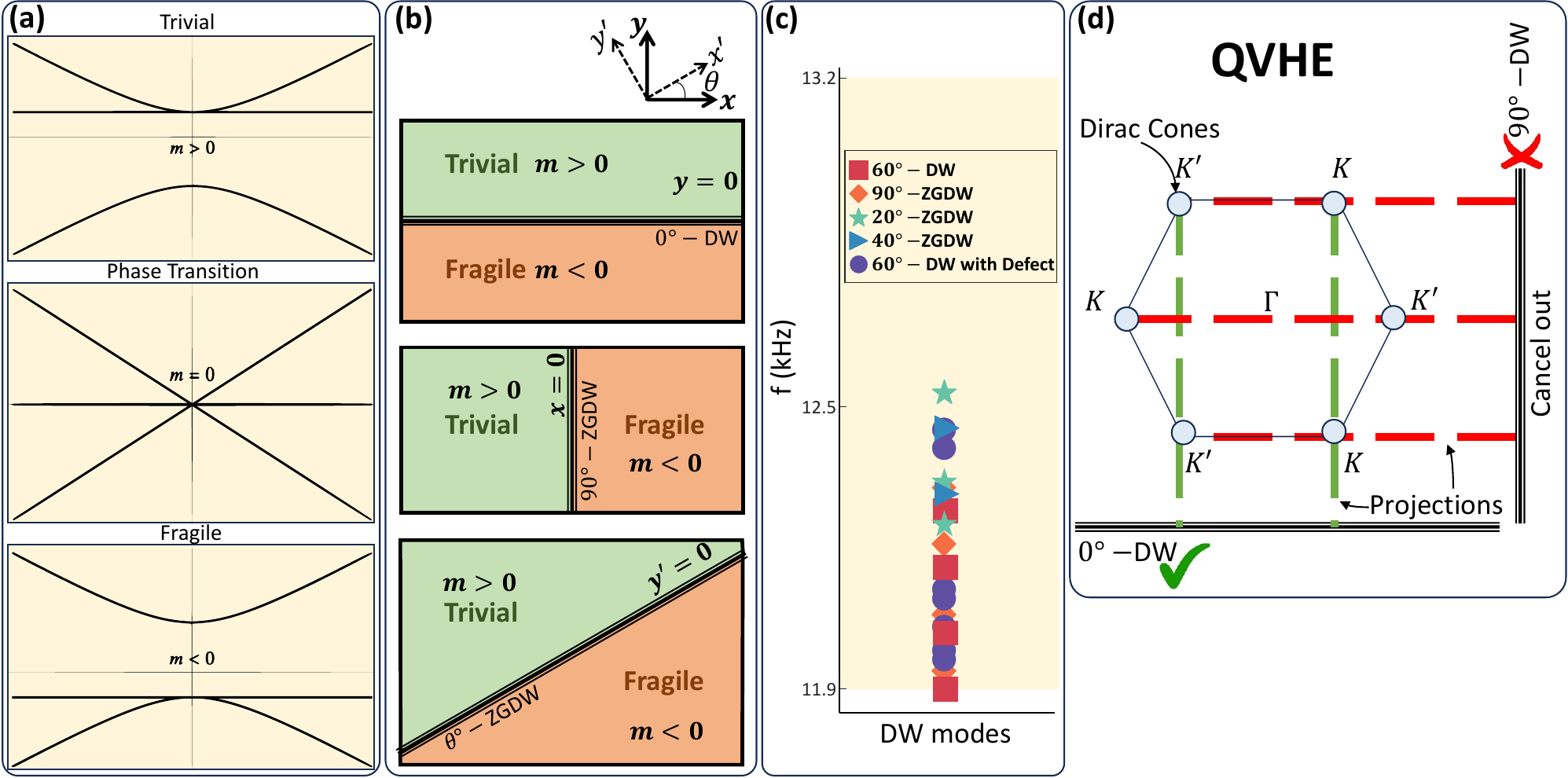}
    \caption{(a) Bands of the $\mathbf{k}\cdot\mathbf{p}$ Hamiltonian in Eq.~\eqref{eq:kpH} for $m>0$ (trivial phase), $m = 0$ (phase transition point), $m<0$ (fragile phase). (b) DWs are described by changing the sign of \enquote{mass} $m$; the upper panel describes the $0^\circ-$DW, the middle panel describes the $90^\circ-$ZGDW, and the lower panel shows the arbitrary oriented DW. (c) Precise DW mode frequencies of the scenarios considered in Fig.~3 of the main text. (d) Difference with the DW from QVHE.}
    \label{fig:jr2}
\end{figure}

Since we are only interested in the behavior near $\Gamma$ point ($\mathbf{k} = \mathbf{0}$), we can expand the matrix $h(\mathbf{k})$ in Taylor series in $\mathbf{k}$. We can write down the most general form of $h(\mathbf{k})$ that satisfies Eq.~\eqref{eq:hSymmetries} up to a linear order in $\mathbf{k}$ 
\begin{equation}
\label{eq:kpH}
\begin{split}
    h(\mathbf{k}) &= \begin{pmatrix}
        \omega_{\Gamma_3}^2 & 0 & 0\\
        0 & \omega_{\Gamma_3}^2 & 0\\
        0 & 0 & \omega_{\Gamma_2}^2
    \end{pmatrix}
    +\begin{pmatrix}
        0 & 0 & - i  a k_y\\
        0 & 0 & i a k_x\\
        i a k_y & -i a k_x & 0
    \end{pmatrix}+ \mathcal{O}(\mathbf{k}^2)\\ 
    &= \frac{2\omega_{\Gamma_3}^2+\omega_{\Gamma_2}^2}{3} \begin{pmatrix}
       1 & 0 & 0\\
        0 & 1 & 0\\
        0 & 0 & 1
    \end{pmatrix} 
    +\begin{pmatrix}
        m & 0 & - i  a k_y\\
        0 & m & i a k_x\\
        i a k_y & -i a k_x & -2m
    \end{pmatrix}+ \mathcal{O}(\mathbf{k}^2) \text{, with } m = \frac{\omega_{\Gamma_3}^2-\omega_{\Gamma_2}^2}{3},
\end{split}
\end{equation}
without going through the actual computation outlined in Eq.~\eqref{eq:Hprojection}. (In the above equation, $a$ is a real constant.) This small $\mathbf{k}$ expansion of the projected Hamiltonian $h(\mathbf{k})$ is known as $\mathbf{k}\cdot\mathbf{p}$ Hamiltonian~\cite{dresselhaus_et_all_group_2007}. In the above expression for the $\mathbf{k}\cdot\mathbf{p}$ Hamiltonian, the term proportional to the identity matrix $\mathds{1}$ shifts the eigenvalues just by a constant amount, but does not affect the structure of the bands or the topology, so we drop it from now on; this shifts the BG between the $\Gamma_3$ and $\Gamma_2$ bands to around $\omega^2 = 0$. The eigenvalues of $h(\mathbf{k})$ are $2m$, $\pm\sqrt{9m^2 + a^2 k^2}-m$ (where we define $k^2 = k_x^2+k_y^2$). Note that the eigenvalues are isotropic in $\mathbf{k}$ due to $\hat{C}_3$ symmetry. These eigenvalues are plotted in Fig.~\ref{fig:jr2}(a) for small $k$ for $m>0$, $m = 0$, and $m<0$. Clearly, for $m>0$, the bands look like bands $3$-$5$ in Fig.~1(c) of the main text (trivial phase), whereas for $m<0$, they resemble bands $3, 4, 7$ in Fig.~1(d) of the main text (fragile phase). The band inversion between $\Gamma_3$ and $\Gamma_2$ from trivial to fragile phase is simply captured by the \enquote{mass} $m$ going from $m>0$ to $m<0$, with $m = 0$ marking the phase transition point. This also justifies why we only keep up to a linear order in $\mathbf{k}$ in the $\mathbf{k}\cdot\mathbf{p}$; the order $\mathbf{k}^2$ terms only change the curvature of the bands, but do not change the essential physics of the band inversion. Another important point to note here is that in principle, one would still have to go through the calculation in Eq.~\eqref{eq:Hprojection} explicitly to obtain the value of $a$; however, below we show that the existence of DW mode does not depend on the value of $a$, only its form and decay length depends on the specific value of $a$. Hence, to prove the existence of the DW mode, the value of $a$ is not important.

\subsection{JR analysis of the DW mode}
Using the $\mathbf{k}\cdot\mathbf{p}$ Hamiltonian derived in the previous section, we show the existence of the DW mode, $60^\circ-$DW, $90^\circ-$ZGDW as well as any generic angle $\theta$ below.
\subsubsection{$60^\circ$( or equivalent $0^\circ$ and $120^\circ$) DW}
The $60^\circ-$DW we consider in the main text is equivalent to a DW that is horizontal (parallel to $x$-axis), due to $\hat{C}_3$ symmetry. In the analysis below, we first use a horizontal DW for simplicity of the calculation, then rotate it by $60^\circ$ to show the plots we obtain using JR analysis. 

Within effective theory, a DW along $y = 0$ between the two phases can be created by changing the sign of the \enquote{mass} $m$ at $y = 0$, as shown in Fig.~\ref{fig:jr2}(b) upper panel; this would describe a DW with fragile (trivial) phase below (above) $y = 0$. Creating a DW (making $m$ dependent on $y$) breaks translation symmetry in $y$-direction; hence we replace $k_y \rightarrow -i \partial_y$ in the Hamiltonian
\begin{equation}
    h(k_x,y) = \begin{pmatrix}
        m(y) & 0 & -a \partial_y\\
        0 & m(y) & i a k_x\\
         a \partial_y & -i a k_x & -2m(y)
    \end{pmatrix} = -i a \lambda_3 \partial_y -a k_x \lambda_7 + \sqrt{3}m(y) \lambda_8,
\end{equation}
where $\lambda_i$ are the well-known Gell-Mann matrices with
\begin{equation}
    \lambda_3 = \begin{pmatrix}
        0 & 0 & -i\\
        0 & 0 & 0\\
        i & 0 & 0
    \end{pmatrix}, \lambda_7 = \begin{pmatrix}
        0 & 0 & 0\\
        0 & 0 & -i\\
        0 & i & 0
    \end{pmatrix}, \lambda_8 = \frac{1}{\sqrt{3}}\begin{pmatrix}
        1 & 0 & 0\\
        0 & 1 & 0\\
        0 & 0 & -2
    \end{pmatrix}.
\end{equation}
\begin{figure}[h!]
    \centering
    \includegraphics[width = 0.7\textwidth]{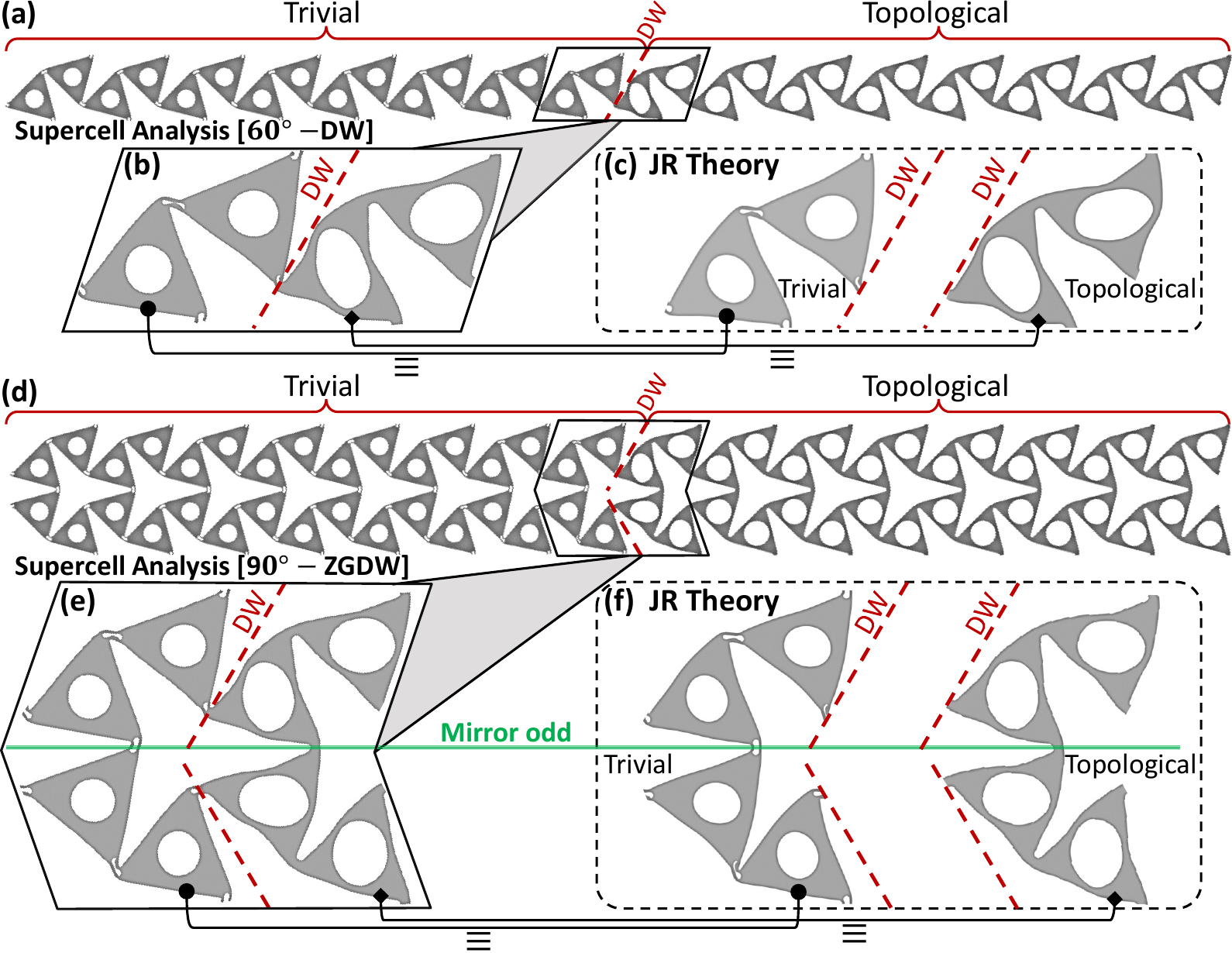}
    \caption{Comparison between DW modes obtained from finite element analysis and JR analysis for (a-c) $60^\circ-$DW and (d-f) $90^\circ-$ZGDW. Panels (a) and (d) are the DW modes from finite elements (same as the ones in Figs.~2(c-d) of the main text). Panels (b) and (e) are respectively the zoomed in representations of (a) and (d) at the DW. Panels (c) and (f) are obtained from JR analysis showing a close agreement with (b) and (e).}
    \label{fig:jr}
\end{figure}

We want an eigenstate of $h(k_x,y)$ in the bulk gap. We start by setting $k_x = 0$, and find the effect of $k_x$ later. Then, we want a three component vector $\boldsymbol{\Psi}(y) = \{\psi_1(y),\psi_2(y),\psi_3(y)\}$, which satisfies $h(k_x =0,y)\boldsymbol{\Psi}(y) = \mathbf{0}$. Notice the following
\begin{equation}
\begin{split}
    &h(k_x =0,y)\Psi(y) = \mathbf{0}\\
    \Rightarrow& (-i a \lambda_3 \partial_y  + \sqrt{3}m(y) \lambda_8)\Psi(y) = \mathbf{0}\\ 
    \Rightarrow& \lambda_8^{-1} (-i a \lambda_3 \partial_y  + \sqrt{3}m(y) \lambda_8)\Psi(y) = \mathbf{0}\\
    \Rightarrow& 
    \begin{pmatrix}
        m(y) & 0 & -a\partial_y\\
        0 & m(y) & 0\\
        -\frac{a}{2}\partial_y & 0 & m(y)
    \end{pmatrix}\boldsymbol{\Psi}(y) = \mathbf{0}.
\end{split}
\end{equation}
For $a = \pm |a|$, choosing $\boldsymbol{\Psi}(y) = f(y)\{1,0,\mp \frac{1}{\sqrt{2}}\}$ simplifies the vector equation above into a scalar equation
\begin{equation}
    (-\frac{a}{\sqrt{2}}\partial_y \mp m(y)) f(y) = 0\Rightarrow f(y) = c_0 e^{\mp\frac{\sqrt{2}}{a}\int_0^y dy'm(y') } = c_0 e^{-\frac{\sqrt{2}}{|a|}\int_0^y dy'm(y') } \Rightarrow \boldsymbol{\Psi}(y) = c_0 e^{-\frac{\sqrt{2}}{|a|}\int_0^y dy'm(y') } \begin{pmatrix}
        1 \\ 0 \\ \mp \frac{1}{\sqrt{2}}
    \end{pmatrix}.
\end{equation}
Hence, we find a solution to $h(k_x =0,y)\boldsymbol{\Psi}(y) = \mathbf{0}$ that is exponentially localized at the DW (regardless of the value of $a$). Then recalling the fact that $h(\mathbf{k})$ is written in the basis $\{\tilde{\mathbf{u}}_1^{(3)}(\mathbf{r}), \tilde{\mathbf{u}}_2^{(3)}(\mathbf{r}),\tilde{\mathbf{u}}^{(2)}(\mathbf{r})\}$, the DW mode in real space has the form $\mathbf{u}_\text{DW}(k_x = 0,\mathbf{r}) = c_0 e^{-\frac{\sqrt{2}}{|a|}\int_0^y dy'm(y') }(\tilde{\mathbf{u}}_1^{(3)}(\mathbf{r}) \mp \frac{1}{\sqrt{2}}\tilde{\mathbf{u}}^{(2)}(\mathbf{r}))$ for $a = \pm |a|$. Now we can use the eigenfunctions correspond to $\Gamma_3$ and $\Gamma_2$ representations obtained from finite element analysis of a single unit cell (shown in Fig.~\ref{fig:symmod}) to construct $\mathbf{u}_\text{DW}(k_x = 0,\mathbf{r})$. In Figs.~\ref{fig:jr}(b-c), we compare the DW mode ($k_x = 0$) obtained from finite element simulation with the mode obtained from JR analysis above. Note that since we do not know the value of $a$, we are plotting $\tilde{\mathbf{u}}_1^{(3)}(\mathbf{r}) \mp \frac{1}{\sqrt{2}}\tilde{\mathbf{u}}^{(2)}(\mathbf{r})$ in Fig.~\ref{fig:jr}(c). Furthermore, since the expression $\tilde{\mathbf{u}}_1^{(3)}(\mathbf{r}) \mp \frac{1}{\sqrt{2}}\tilde{\mathbf{u}}^{(2)}(\mathbf{r})$ is dependent on the sign of $a$ (which is unknown a priori), we check both the functions, and in Fig.~\ref{fig:jr}(c) we plot the one (that corresponds to $+$ sign of $a$) that is similar to the DW mode obtained from finite elements. Clearly, the mode shape in Fig.~\ref{fig:jr}(c) is very close to Fig.~\ref{fig:jr}(b) which confirms that the DW modes from the simulation are indeed the same as the JR mode, in other words the existence of the DW mode is due to the band inversion between the trivial and fragile phases.

Next, we comment on nonzero $k_x$ DW mode. Due to the continuity of eigenvalues as a function of $k_x$, for a small nonzero $k_x$ there would still be a DW mode since the eigenvalue cannot change abruptly for a small $k_x$. We can try to understand the dependence of the eigenvalue as a function of $k_x$ perturbatively for small $|k_x|$. To the linear order in $k_x$, the change in the eigenvalue from $k_x = 0$ eigenvalue is
\begin{equation}
    E_\text{DW}(k_x) = \langle \boldsymbol{\Psi}(y)|-a k_x \lambda_7+ \mathcal{O}(k_x^2) | \boldsymbol{\Psi}(y) \rangle = 0 + \mathcal{O}(k_x^2).
\end{equation}
Therefore, the dispersion of the DW mode near $k_x = 0$ is at least of quadratic order. This has to be true since if there was a linear dispersion at $k_x = 0$ for a single DW mode, that would make the DW mode, chiral (like Chern insulators), which would break the time reversal symmetry.

\subsubsection{$90^\circ$( or equivalent $30^\circ$ and $150^\circ$) ZGDW}
The $90^\circ-$ZGDW is along the $y$ direction, we can take it to be at $x = 0$. Within effective theory, a DW along $x = 0$ between the two phases can be created by changing the sign of the \enquote{mass} $m$ at $x = 0$, as shown in Fig.~\ref{fig:jr2}(b) middle panel; this would describe a DW with fragile (trivial) phase at $x>0$ ($x<0$). The rest of the analysis is the same as that for the $60^\circ-$DW. Creating a DW (making $m$ dependent on $x$) breaks translation symmetry in $x$-direction; hence we replace $k_x \rightarrow -i \partial_x$ in the Hamiltonian
\begin{equation}
    h(x,k_y) = \begin{pmatrix}
        m(x) & 0 & -iak_y\\
        0 & m(x) & a \partial_x\\
         iak_y & -a \partial_x & -2m(x)
    \end{pmatrix} =  a \lambda_3 k_y +ia  \lambda_7 \partial_x + \sqrt{3}m(x) \lambda_8.
\end{equation}
We want an eigenstate of $h(x,k_y)$ in the bulk gap. We start by setting $k_y = 0$, and find the effect of $k_y$ later. Then, we want a three component vector $\boldsymbol{\Psi}(x) = \{\psi_1(x),\psi_2(x),\psi_3(x)\}$, which satisfies $h(x,k_y =0)\boldsymbol{\Psi}(x) = \mathbf{0}$. Notice the following
\begin{equation}
\begin{split}
    &h(x,k_y=0)\Psi(x) = \mathbf{0}\\
    \Rightarrow& (i a \lambda_7 \partial_x  + \sqrt{3}m(x) \lambda_8)\Psi(x) = \mathbf{0}\\ 
    \Rightarrow& \lambda_8^{-1} (i a \lambda_7 \partial_x  + \sqrt{3}m(x) \lambda_8)\Psi(x) = \mathbf{0}\\
    \Rightarrow& 
    \begin{pmatrix}
        m(x) & 0 & 0\\
        0 & m(x) & a \partial_x\\
        0 & \frac{a}{2} \partial_x & m(x)
    \end{pmatrix}\boldsymbol{\Psi}(x) = \mathbf{0}.
\end{split}
\end{equation}
For $a = \pm |a|$, choosing $\boldsymbol{\Psi}(x) = f(x)\{0,1,\mp \frac{1}{\sqrt{2}}\}$ simplifies the vector equation above into a scalar equation
\begin{equation}
    (\frac{a}{\sqrt{2}}\partial_x \mp m(x)) f(x) = 0\Rightarrow f(x) = c_0 e^{\pm\frac{\sqrt{2}}{a}\int_0^x dx'm(x') } = c_0 e^{\frac{\sqrt{2}}{|a|}\int_0^x dx'm(x') } \Rightarrow \boldsymbol{\Psi}(x) = c_0 e^{\frac{\sqrt{2}}{|a|}\int_0^x dx'm(x') } \begin{pmatrix}
        0 \\ 1 \\ \mp \frac{1}{\sqrt{2}}
    \end{pmatrix}.
\end{equation}
Hence, we find a solution to $h(x,k_y=0)\boldsymbol{\Psi}(x) = \mathbf{0}$ that is exponentially localized at the DW (regardless of the value of $a$). Then recalling the fact that $h(\mathbf{k})$ is written in the basis $\{\tilde{\mathbf{u}}_1^{(3)}(\mathbf{r}), \tilde{\mathbf{u}}_2^{(3)}(\mathbf{r}),\tilde{\mathbf{u}}^{(2)}(\mathbf{r})\}$, the DW mode in real space has the form $\mathbf{u}_\text{DW}(k_y = 0,\mathbf{r}) = c_0 e^{\frac{\sqrt{2}}{|a|}\int_0^x dx'm(x') }(\tilde{\mathbf{u}}_2^{(3)}(\mathbf{r}) \mp \frac{1}{\sqrt{2}}\tilde{\mathbf{u}}^{(2)}(\mathbf{r}))$ for $a = \pm |a|$. Now, we can use the eigenfunctions correspond to $\Gamma_3$ and $\Gamma_2$ representations obtained from finite element analysis of a single unit cell (shown in Fig.~\ref{fig:symmod}) to construct $\mathbf{u}_\text{DW}(k_x = 0,\mathbf{r})$. In Figs.~\ref{fig:jr}(e-f), we compare the DW mode ($k_y = 0$) obtained from finite element simulation with the mode obtained from JR analysis above. Note that since we do not know the value of $a$, we plot $\tilde{\mathbf{u}}_2^{(3)}(\mathbf{r}) \mp \frac{1}{\sqrt{2}}\tilde{\mathbf{u}}^{(2)}(\mathbf{r})$ in Fig.~\ref{fig:jr}(f). Furthermore, since the expression $\tilde{\mathbf{u}}_2^{(3)}(\mathbf{r}) \mp \frac{1}{\sqrt{2}}\tilde{\mathbf{u}}^{(2)}(\mathbf{r})$ depends on the sign of $a$ (which is unknown a priori), we check both the functions, and in Fig.~\ref{fig:jr}(f) we plot the one (that corresponds to $+$ sign of $a$) that is similar to the DW mode obtained from finite elements. Clearly, the mode shape in Fig.~\ref{fig:jr}(f) is very close to Fig.~\ref{fig:jr}(e) which confirms that the DW modes from the simulation are indeed the same as the JR mode. In other words, the existence of the DW mode is due to the band inversion between the trivial and fragile phases. It is worthwhile to mention that since both $\tilde{\mathbf{u}}_2^{(3)}$ and $\tilde{\mathbf{u}}^{(2)}$ are odd under the mirror $\hat{M}_y$, the DW mode from JR analysis is odd under the mirror $\hat{M}_y$ (note that the $90^\circ-$ZGDW is mirror $\hat{M}_y$ symmetric as shown in Figs.~\ref{fig:jr}(e-f), unlike the $60^\circ-$DW which breaks the mirror symmetry); this prediction from JR analysis also matches with the finite element analysis result.

Next we comment on nonzero $k_y$ DW mode. Due to the continuity of eigenvalues as a function of $k_y$, for a small nonzero $k_y$ there would still be a DW mode since the eigenvalue cannot change abruptly for small $k_y$. We can try to understand the dependence of the eigenvalue as a function of $k_y$ perturbatively for small $|k_y|$. To the linear order in $k_y$, the change in the eigenvalue from $k_y = 0$ eigenvalue is
\begin{equation}
    E_\text{DW}(k_y) = \langle \boldsymbol{\Psi}(x)|a k_y \lambda_3+ \mathcal{O}(k_y^2)| \boldsymbol{\Psi}(x)\rangle  = 0 + \mathcal{O}(k_y^2).
\end{equation}
Therefore, the dispersion of the DW mode near $k_y = 0$ is at least of quadratic order. This has to be true since if there was a linear dispersion at $k_y = 0$ for a single DW mode, that would make the DW mode, chiral (like Chern insulators), which would break the time reversal symmetry.

\subsubsection{Generalization to an arbitrary oriented DW}
Notice that the derivations for above two angles are very similar. Here we show that a similar analysis can be done for a DW at any angle $\theta$ along the line $y = x\tan\theta$. To do this, we rotate the coordinate system to new coordinates $(x',y')$ such that $x'$ is along the line $y = x\tan\theta$. In other words, we choose $x=x'\cos\theta - y'\sin\theta$ and $y=x'\sin\theta + y'\cos\theta$. In this new coordinate system the DW is along the line $y'=0$. Accordingly $h(\mathbf{k})$ becomes
\begin{equation}
    h(\mathbf{k})= \frac{2\omega_{\Gamma_3}^2+\omega_{\Gamma_2}^2}{3} \begin{pmatrix}
       1 & 0 & 0\\
        0 & 1 & 0\\
        0 & 0 & 1
    \end{pmatrix} 
    +\begin{pmatrix}
        m & 0 & - i  a (k_{x'}\sin\theta + k_{y'}\cos\theta)\\
        0 & m & i a (k_{x'}\cos\theta - k_{y'}\sin\theta)\\
        i a (k_{x'}\sin\theta + k_{y'}\cos\theta) & -i a (k_{x'}\cos\theta - k_{y'}\sin\theta) & -2m
    \end{pmatrix}+ \mathcal{O}(\mathbf{k}^2)
\end{equation}
The rest of the analysis follows the same arguments as that for the $60^\circ$-DW. Within effective theory, a DW along $y' = 0$ between the two phases can be created by changing the sign of the \enquote{mass} $m$ at $y' = 0$, as shown in Fig.~\ref{fig:jr2}(b) lower panel; this would describe a DW with fragile (trivial) phase below (above) $y' = 0$. Creating a DW (making $m$ dependent on $y'$) breaks translation symmetry in $y'$-direction; hence we replace $k_{y'} \rightarrow -i \partial_{y'}$ in the Hamiltonian (and omitting the trivial term proportional to identity)
\begin{equation}
    h(k_{x'},y') = \begin{pmatrix}
        m(y') & 0 & - i  a k_{x'}\sin\theta -a\cos\theta\partial_{y'}\\
     0 & m(y') & i a k_{x'}\cos\theta - a\sin\theta\partial_{y'}\\
          i  a k_{x'}\sin\theta +a\cos\theta\partial_{y'} & -i a k_{x'}\cos\theta + a\sin\theta\partial_{y'} & -2m(y')
    \end{pmatrix}.
\end{equation}
We want an eigenstate of $h(k_{x'},y')$ in the bulk gap. We start by setting $k_{x'} = 0$, and find the effect of $k_{x'}$ later. Then, we want a three component vector $\boldsymbol{\Psi}(y') = \{\psi_1(y'),\psi_2(y'),\psi_3(y')\}$, which satisfies $h(k_{x'} =0,y')\boldsymbol{\Psi}(y') = \mathbf{0}$. Notice the following
\begin{equation}
\begin{split}
    &h(k_{x'} =0,y')\Psi(y') = \mathbf{0}\\
    \Rightarrow& \lambda_8^{-1} h(k_{x'} =0,y')\Psi(y') = \mathbf{0}\\
    \Rightarrow& 
    \begin{pmatrix}
        m(y') & 0 & -a\cos\theta\partial_{y'}\\
        0 & m(y') & -a\sin\theta\partial_{y'}\\
        -\frac{a}{2}\cos\theta\partial_{y'} & -\frac{a}{2}\sin\theta\partial_{y'} & m(y')
    \end{pmatrix}\boldsymbol{\Psi}(y') = \mathbf{0}.
\end{split}
\end{equation}
For $a = \pm |a|$, choosing $\boldsymbol{\Psi}(y') = f(y')\{\cos\theta,\sin\theta,\mp \frac{1}{\sqrt{2}}\}$ simplifies the vector equation above into a scalar equation
\begin{equation}
    (-\frac{a}{\sqrt{2}}\partial_{y'} \mp m(y')) f(y') = 0\Rightarrow f(y') = c_0 e^{\mp\frac{\sqrt{2}}{a}\int_0^{y'} dy''m(y'') } = c_0 e^{-\frac{\sqrt{2}}{|a|}\int_0^{y'} dy''m(y'') } \Rightarrow \boldsymbol{\Psi}(y') = c_0 e^{-\frac{\sqrt{2}}{|a|}\int_0^{y'} dy''m(y'') } \begin{pmatrix}
        \cos\theta\\\sin\theta \\ \mp \frac{1}{\sqrt{2}}
    \end{pmatrix}.
\end{equation}
Hence, we find a solution to $h(k_{x'} =0,y')\boldsymbol{\Psi}(y') = \mathbf{0}$ that is exponentially localized at the DW (regardless of the value of $a$). For nonzero $k_{x'}$, a similar analysis as the two previous subsections can be done. We also list the precise DW mode frequencies of the scenarios considered in Fig. 3 of the main text in Fig.~\ref{fig:jr2}(c), emphasizing that despite the slight variability between the exact frequencies, they display substantial stability within the identified range for the considered scenarios, even though the sizes of the finite bi-domain configurations differ between cases.

\subsubsection{Difference with Quantum Valley Hall Effect (QVHE)}
Our system is reminiscent of QVHE, where DW modes also arise. However there is an important difference. In QVHE, DW modes cannot appear when the DW is along either of the angles $30^\circ$, $90^\circ$ and $150^\circ$ from one of the lattice vectors. This can be understood as follows. In systems with 3-fold rotation symmetry and broken 2-fold rotation symmetry, the Dirac cones at $K$ and $K'$ points gap out. However, these gapped bands have concentration of Berry curvature at $K$ and $K'$ points since they originate from the Dirac cone which has a topological winding. Furthermore, the Berry curvatures at $K$ and $K'$ are equal in magnitude and opposite in sign due to time reversal symmetry of the system. Now, when we create a DW, the Berry curvature concentrations at $K$ and $K'$ points result in the DW modes. However, when the DW is at the particular angles $30^\circ$, $90^\circ$ and $150^\circ$ from the lattice vectors, $K$ and $K'$ points project onto the same point in the 1D DW Brillouin zone (BZ) as shown in Fig.~\ref{fig:jr2}(d). Since $K$ and $K'$ points have opposite Berry curvatures, the total Berry curvature is zero at the projected point, hence there is no DW mode at these angles. 

The argument above also elucidates why the DW mode in our system is present for any angle. The DW mode in our system originates from a band inversion at the $\Gamma$ point. Unlike $K$ and $K'$ points which are time reversal partners with opposite Berry curvature, there is only one $\Gamma$ point in the BZ. Hence, we do not need to worry about points with opposite Berry curvature projecting to the same point in the 1D DW BZ. This is why the DW mode in our system is present for any angle of DW.

Another way of seeing that the DW originating from $\Gamma$ point band inversion must be omni-directional is the fact that the $\Gamma$ point is isotropic due to the 3-fold rotation symmetry.

%\pa{\subsubsection{Extension to arbitrary oriented DWs}
%In the previous sections, we assert that the bi-domain configuration supports robust DW modes, irrespective of the DW orientation. This property is demonstrated for two specific cases: the straight DW ($60^{\circ}$) and the zigzag DW ($90^{\circ}$). Through full-scale simulations, theoretical considerations supported by JR analysis, and experimental validation for the straight DW case, we substantiate our claim. In this section, our goal is to extend this phenomenon to arbitrary angled DWs. Due to the inherent symmetry of the model, only DW orientations ranging from $0^{\circ}$ to $59^{\circ}$ yield distinct DWs. That being said, the previously mentioned zigzag DW ($90^{\circ}$) is equivalent to DWs oriented at $30^{\circ}$ or $150^{\circ}$. Here, we present two additional possibilities with DW inclinations of $20^{\circ}$ and $40^{\circ}$. Eigenfrequencies within the frequency interval of the shared BG for these two cases are depicted in Fig.~\ref{fig:genDW}, with localized modes specified in the insets. Importantly, our method of constructing different DWs clearly demonstrates that any desired orientation can be achieved through a suitable combination of zigzag forms.}
%\begin{figure}[h!]
%    \centering
 %   \includegraphics[width = 0.9\textwidth]{figS6.pdf}
    %\caption{\pa{Eigenfrequencies in the frequency interval of the shared BG with localized modes, including corner, edge, and DW modes specified in the insets, for the two DW orientations: (a) $20^{\circ}$, and (b) $40^{\circ}$.} }
  %  \label{fig:genDW}
%\end{figure}

\section{Material properties, geometrical details of the specimen and experimental setup}\label{sec.3}
\begin{figure}[h!]
    \centering
    \includegraphics[width = 0.6\textwidth]{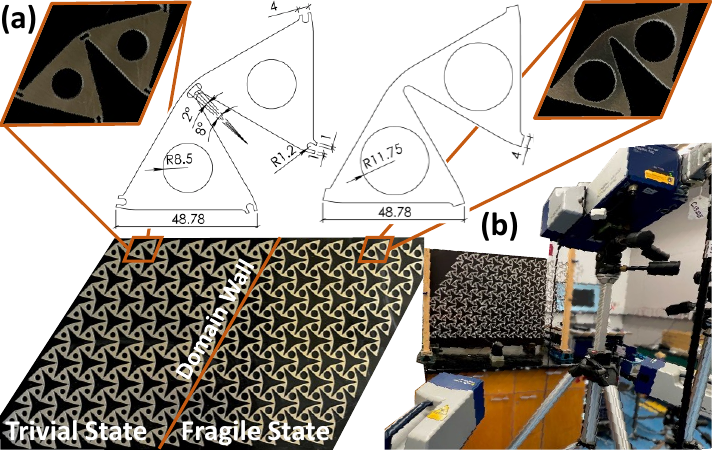}
    \caption{(a) Manufactured specimen featuring trivial and fragile states on the left and right side of the DW, respectively, with details of the unit cells geometry in the insets. All unit lengths are in mm. (b) Experimental setup for 3D SLDV testing of the prototype.}
    \label{fig:det}
\end{figure}

After finding a compatible pair of SSTK unit cells with distinct topological nature, one trivial and the other fragile, we design a finite element model of the stitched domain forming a $60^\circ-$DW. The finite bi-domain configuration consists of two parallelograms of trivial and fragile states stitched together, discretized with a mesh of plane-stress quadrilateral isoparametric elements, with appropriate mesh refinement surrounding the holes. The model is designed using SOLIDWORKS and is exported as a “.SLDPRT” file for the manufacturing process. Fig.~\ref{fig:det}(a) shows the geometry of the manufactured specimen obtained via water-jet cutting from a 2-mm thick sheet of Aluminum, along with the trivial and fragile unit cells. The inset provides further details of the lengths of the sides of the triangles, the hinge thicknesses, central holes radii, and other geometric specifications for the bean-shaped hole  at the hinge of the trivial state. The material properties of Aluminum are: Young’s modulus E = 71 GPa, Poisson’s ratio $\nu$ = 0.33 and density $\rho = 2700$ kg/m$^3$. The experimental setup is depicted in Fig.~\ref{fig:det}(b), which shows the specimen constrained via boundary supports and the heads of the 3D Scanning Laser Doppler Vibrometer (SLDV) employed to measure in-plane velocity. We apply excitation by an electromechanical shaker (Brüel \& Kj\ae r Type 4810) with an amplifier (Brüel \& Kj\ae r Type 2718). In-plane velocity is measured at three scan points per triangle. Retro-reflective tape is applied at the scan points to increase the reflectivity and reduce noise in the data.

%\section{Full-scale steady-state simulation transmissibility curve}
%\begin{figure}[h!]
 %   \centering
  %  \includegraphics[width = 0.9\textwidth]{figS8.pdf}
   % \caption{Transmissibility curve obtained from simulations, with mid-frequency BG region highlighted in pink and color-coded to distinguish different types of modes, including bulk, DW, and edge modes. Three randomly selected eigenfunctions corresponding to bulk, DW, and edge states and located at 7.13 kHz, 12.2 kHz, and 13.2 kHz, respectively, are also illustrated.}
    %\label{fig:trans}
%\end{figure}
%We perform full-scale steady-state simulation with sustained harmonic excitation applied at the bottom edge of the model, at a point marking the end of the DW, using a force polarized perpendicularly to the DW. To create a transmissibility curve, we normalize the magnitude of displacement sampled at at a point located midway along the DW by the magnitude of displacement of the loaded point. The curve, depicted in Fig.~\ref{fig:trans}, shows an attenuation region (marked in pink) in a frequency interval corresponding to the shared BG from the band diagram. The distinct modes recognized in the response, i.e., bulk, DW-bound, and edge modes, are color-coded for clarity. We also provide illustrations of three eigenstates, each corresponding to one of the above-mentioned mode types (in the plots, we mark the input and output points with a star and a dot, respectively). 
\section{Full-Scale Steady-State Simulation Transmissibility Curve}\label{sec.4}

\begin{figure}[h!]
    \centering
    \includegraphics[width=0.8\textwidth]{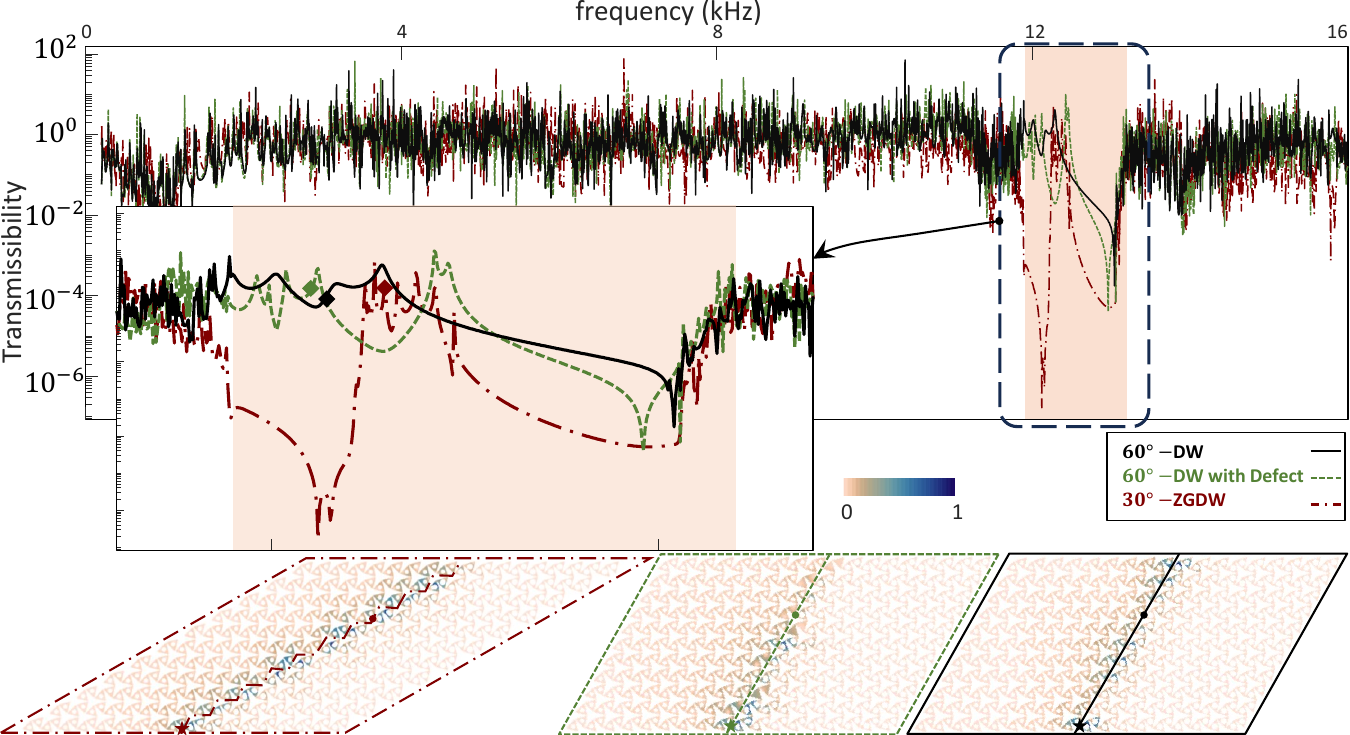}
    \caption{Transmissibility curve obtained from simulations, highlighting the mid-frequency BG region in pink for three cases: $60^\circ-$DW, $60^\circ-$DW with defects, and $30^\circ-$ZGDW. The inset zooms in on transmissibility within the BG frequency interval, with selected eigenfields marked by diamond symbols corresponding to DW modes in each configuration.}
    \label{fig:trans}
\end{figure}

We conduct full-scale steady-state simulations under sustained harmonic excitation for three cases: $60^\circ-$DW, $60^\circ-$DW with defects, and $30^\circ-$ZGDW. To ensure a fair comparison, we modify the configuration of the $90^\circ-$ZGDW presented in the main text, adjusting it to have the same number of unit cells in both the fragile and trivial parts as the other two configurations. Additionally, we consider a DW inclination of $30^{\circ}$, which is equivalent to the $90^\circ-$ZGDW presented in the main text as justified in earlier sections, for a more similar finite domain configuration. In each configuration, harmonic excitation is applied at the bottom edge of the model, at the end of the DW, using a force polarized perpendicular to the DW. To create a transmissibility curve, we normalize the displacement magnitude sampled at a point midway along the DW by the magnitude of displacement at the loaded point. The resulting curve, depicted in Fig.~\ref{fig:trans}, exhibits an attenuation region (highlighted in pink) within a frequency interval corresponding to the shared BG from the band diagram. We also provide illustrations of three eigenstates, each corresponding to one of the DW modes in each configuration. In the plots, we mark the input and output points with a star and a dot, respectively. The curves reveal slight differences in the transmission spectrum between the cases, particularly in the BG region. However, in each case, there are frequencies where a high transmission rate is observed along the DW, evident in the curve and mode shapes provided.
\end{document}